\pdfoutput=1 

\documentclass[sigconf]{acmart}

\copyrightyear{2026}
\acmYear{2026}
\setcopyright{cc}
\setcctype{by}
\acmConference[UIST '26]{The 39th Annual ACM Symposium on User Interface Software and Technology}{November 02--05, 2026}{Detroit, MI, USA}
\acmBooktitle{The 39th Annual ACM Symposium on User Interface Software and Technology (UIST '26), November 02--05, 2026, Detroit, MI, USA}
\acmDOI{10.1145/3830398.3830522}
\acmISBN{979-8-4007-2856-3/2026/11}

\usepackage{acmart-taps}
\usepackage{framed}

\usepackage{latexml}
\newcommand{\inlineicon}[2]{\iflatexml\else\includegraphics[height=#1]{#2}\fi}

\usepackage{enumitem}
\aptLtoXcmd{}{\setlist[itemize]{leftmargin=*}
\setlist[enumerate]{leftmargin=*,label=\arabic*.}}

\usepackage{xcolor}

\usepackage{xspace}

\newcommand{\ie}{\emph{i.e.{\xspace}}}
\newcommand{\eg}{\emph{e.g.{\xspace}}}

\definecolor{GREEN}{rgb}{0.0,0.6,0.0}
\definecolor{BLUE}{rgb}{0.0,0.2,0.7}
\definecolor{GOLD}{rgb}{0.5,0.5,0.0}
\definecolor{CYAN}{rgb}{0.0,0.5,0.5}
\definecolor{PURPLE}{rgb}{0.5,0.0,0.5}

\definecolor{RED}{rgb}{0.7,0.0,0.0}
\definecolor{GRAY}{gray}{0.5}

\definecolor{LIGHTGRAY}{HTML}{DDDDDD}

\definecolor{REVISIONYELLOW}{HTML}{FFF176} 
\definecolor{REVISIONGREEN}{HTML}{427a11}
\definecolor{REVISIONGOLD}{HTML}{806b06}
\definecolor{REVISIONBLUE}{HTML}{0000CC}
\definecolor{REVISIONCYAN}{HTML}{80D8FF}   

\makeatletter
\aptLtoXcmd{}{\renewcommand{\@subsubsecfont}{\bfseries}}
\makeatother

\usepackage{fancybox}

\usepackage{multirow}

\definecolor{dg1bg}{HTML}{D6E8F5}
\definecolor{dg1text}{HTML}{1A4F72}
\definecolor{dg2bg}{HTML}{FFF3CD}
\definecolor{dg2text}{HTML}{7A5C00}
\definecolor{dg3bg}{HTML}{D9EDD4}
\definecolor{dg3text}{HTML}{2D5E25}
\definecolor{mainpeach}{HTML}{EE9E7A}
\definecolor{deeppeach}{HTML}{D4714A}
\definecolor{darkpeach}{HTML}{9C4A2A}
\definecolor{mainpurple}{HTML}{9D94E0}
\definecolor{deeppurple}{HTML}{6E64C4}
\definecolor{darkpurple}{HTML}{4A4191}

\definecolor{ie}{HTML}{4A148C}
\definecolor{ce}{HTML}{C65BD8}
\definecolor{tb}{HTML}{CE93D8}
\definecolor{kb}{HTML}{BF360C}
\definecolor{llm}{HTML}{FF7043}
\definecolor{gs}{HTML}{FFAB91}
\definecolor{ui}{HTML}{BDBDBD}

\newcommand{\designconcept}[4]{%
  {\setlength{\fboxsep}{1pt}%
  \colorbox{#3}{%
    \textbf{\textcolor{#4}{#2}}%
  }}%
}

\newsavebox{\tcardtitle}
\newsavebox{\tcardall}

\newlength{\tcardhpad}
\newlength{\tcardtpad}
\newlength{\tcardbpad}

\newcommand{\taskcard}[3]{%
  \begingroup
  \cornersize*{4pt}%
  \begin{lrbox}{\tcardtitle}%
    \begin{minipage}{\dimexpr\linewidth-2pt\relax}%
      \setlength{\parindent}{0pt}%
      \setlength{\parskip}{0pt}%
      \vspace{\tcardtpad}%
      \leftskip=\tcardhpad\relax
      \rightskip=\tcardhpad\relax
      \small\bfseries\color{white}#2\par
      \vspace{\tcardtpad}%
    \end{minipage}%
  \end{lrbox}%
  \begin{lrbox}{\tcardall}%
    \begin{minipage}{\dimexpr\linewidth-2pt\relax}%
      \setlength{\parindent}{0pt}%
      \setlength{\parskip}{0pt}%
      \noindent{\setlength{\fboxsep}{0pt}\colorbox{#1}{\usebox{\tcardtitle}}}\par
      \vspace{\tcardbpad}%
      \leftskip=\tcardhpad\relax
      \rightskip=\tcardhpad\relax
      \small #3\par
      \vspace{\tcardbpad}%
    \end{minipage}%
  \end{lrbox}%
  {\setlength{\fboxsep}{0pt}\color{#1}\Ovalbox{\usebox{\tcardall}}}%
  \endgroup
}

\makeatletter
\g@addto@macro\normalsize{%
  \setlength\abovedisplayshortskip{-9pt}
  \setlength\belowdisplayshortskip{3pt}
}
\makeatother

\usepackage{color} 

\newcommand{\bstart}[1]{\vspace{1mm} \noindent{\textbf{#1.}}}

\begin{document}

\tolerance=400 

\title[InsightToast]{\textit{InsightToast}: Proactive Information Retrieval \& Glanceable Visualization in the Side Channel of Data-Rich Meetings}

\author{Mohammad Abolnejadian}
\affiliation{%
  \department{School of Computer Science}
  \institution{University of Waterloo}
  \city{Waterloo}
  \state{Ontario}
  \country{Canada}}
\email{mabolnej@uwaterloo.ca}
\orcid{0000-0003-2028-9270}

\author{Matthew Brehmer}
\affiliation{%
  \department{School of Computer Science}
  \institution{University of Waterloo}
  \city{Waterloo}
  \state{Ontario}
  \country{Canada}
}
\email{mbrehmer@uwaterloo.ca}
\orcid{0000-0001-5524-2291}

\begin{abstract}


Missing institutional context during meetings can impede effective participation. Retrieving relevant information, often scattered across heterogeneous internal and external sources, requires costly task-switching that disrupts both individual focus and collective conversational flow, particularly detrimental during cognitively demanding tasks such as decision-making. We introduce \textsc{InsightToast}, a mixed-initiative application that monitors verbal discourse in real time, identifies topics and informational needs as they emerge, and proactively retrieves relevant information through a multi-agent large language model (LLM)-based pipeline integrating retrieval-augmented generation (RAG) to produce source-grounded insights as succinct text and glanceable interactive charts, delivered through a peripheral interface as ephemeral toasts in the conversation's side channel. To demonstrate the potential for yielding serendipitous insights, we showcase a usage scenario involving a knowledge base of legislative documents as the meeting's context. We then report on a comparative study ($N=16$), in which participants arrived at informed policy decisions while maintaining natural conversation flow.


\end{abstract}



\begin{teaserfigure}
  \includegraphics[width=\textwidth]{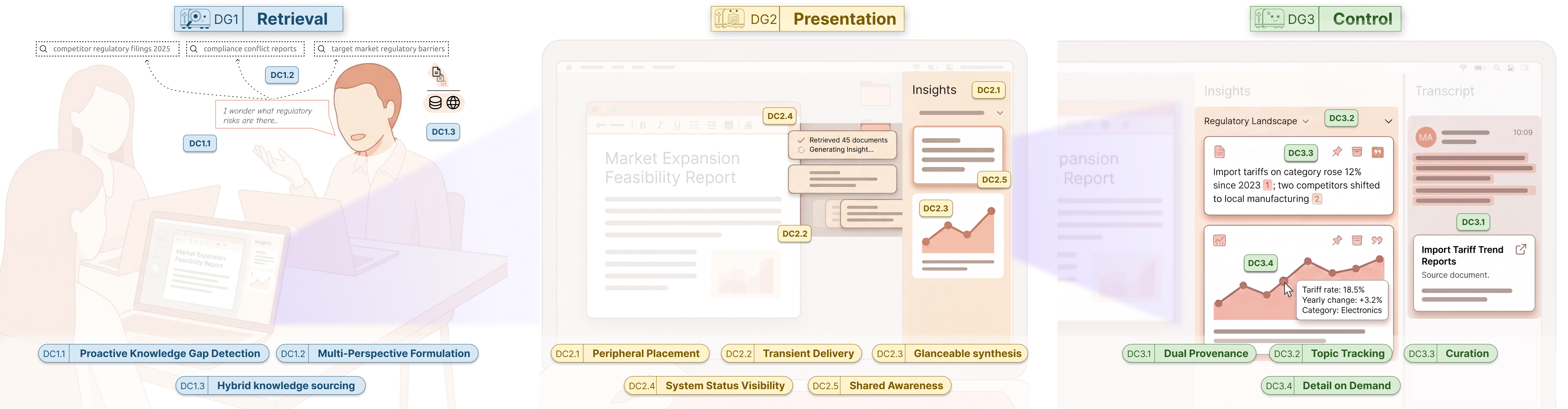}
  \caption{\textsc{InsightToast} is a mixed-initiative interface that lives in the periphery of a meeting, proactively retrieves information as knowledge gaps emerge, and surfaces glanceable, source-traceable, and context-aware insights through ephemeral toasts.}
  \Description{Three side-by-side annotated illustrations of a frame of a co-located business team meeting. The left frame (tagged with Design Goal 1: Retrieval) depicts three people at a table with each having a laptop open in front of them, one with a speech bubble saying "I wonder what regulatory risks are there," which triggers retrieval proactively (tagged with Design Concept 1.1: Proactive Knowledge Gap Detection) by spawning three complementary search queries (tagged with Design Concpept 1.2: Multi-Perspective Formulation) shown above as search bars (Search bar 1: competitor regulatory filings 2025, Search bar 2: compliance conflict reports, Search bar 3: target market regulatory barriers) connected from the speech bubble by arrows. The search bars are connected to two web and knowledgebase icons with flying document icons (tagged with Design Concept 1.3: Hybrid Knowledge Sourcing). The center frame (tagged with Design Goal 2: Presentation) is a zoomed-in view of one of the meeting attendees' laptop screen with a Market Expansion Feasibility Report in the main view, three toast notifications in the top-right corner with the top one reporting retrieval and synthesis progress (tagged with Design Concept 2.4: System Status Visibility) and the other two showing abstract text lines with the one in the bottom having a ghost effect (tagged with Design Concept 2.2: Transient Delivery), an expandable side panel with the word "Insights" being read on top of it on the right (tagged with Design Concept 2.1: Peripheral Placement), and two insight cards depicted in this side panel, with the top one showing a text insight card with glowing borders (tagged with Design Concept 2.5: Shared Awareness) and the bottom one showing a simple line chart (tagged with Design Concept 2.3: Glanceable Synthesis). The right frame (tagged with Design Goal 3: Control) is a zoomed-in view of the side panel that was depicted in the middle frame, along with another secondary side panel opened to its right. In the main side panel, there is an insight card with three action icons of pin, archive, and source on its top right (tagged with Design Concpet 3.3: Curation) and a chart insight showing a line chart with a cursor hovering on top of one of its data points, displaying a tooltip that reads "tarrif rates, yearly change, and category" (tagged with Design Concept 3.4: Detail on Demand), both organized under a "Regulatory Landscape" topic heading (tagged with Design Concept 3.2: Topic Tracking). The secondary side panel, which is to the right of the main one, shows an abstract text graphic that is highlighted with an "MA" avatar in front of it, and below it a document card that reads "Import Tariff Trend Reports, Source document.) with an open in new tab icon on its top-right corner (tagged with Design Concept 3.1: Dual Provenance).}
  \label{fig:design-space}
  \label{fig:teaser}
\end{teaserfigure}

\maketitle





\section{Introduction}
\label{sec:introduction}

Every meeting has context. 
Whenever people meet to make decisions, plan, or ideate together, context includes institutional knowledge, the agendas and transcripts of prior meetings, and data relevant to the topic at hand~\cite{Brehmer2026-jo}.
However, apart from prepared materials such as slide presentations or agendas that establish context, retrieving and sharing \textit{additional} context and data during a meeting can be highly disruptive {to the flow of the conversation}.
Disruptions arise from application-switching to search interfaces and by side channel conversations in which meeting participants share missing context. 
Irrespective of how missing context is sought, people require the right keywords to search or ask about, and this poses a challenge when the meeting's domain is unfamiliar and they lack the terminology (and attentional resources) to articulate what information they need~\cite{Marchionini2006-iv}. 
On the other hand, a conscious avoidance of these disruptions can lead participants to either defer information retrieval until after the meeting or to act on whatever information is available during the meeting, a \textit{What You See Is All There Is} bias~\cite{Kahneman2011-en} which risks making under-informed decisions.

With the advent of large language models (LLMs), many videoconferencing platforms (\eg~Microsoft Teams \cite{MicrosoftTeamsAI2026} and Google Meet \cite{WorkspaceUnknown-rn}) now offer meeting support through transcription and retrospective summaries, as well as in-meeting assistance via conversational question answering (\eg~Zoom's AI Companion \cite{ZoomCustomAI2026}).
However, these tools remain reactive, requiring conscious initiation and thus perpetuating the interruption costs described above. 
We instead posit that information retrieval during synchronous human-to-human conversations should be as effortless and ambient as possible, operating in the background without demanding conscious initiation and delivering results unobtrusively.
{In doing so, a communication platform can enhance meeting effectiveness by establishing common ground among the group through addressing participants' individual information needs as they emerge.~\cite{Kim2016-hl}}
While prior work has explored proactive {retrieval~\cite{Rhodes2000-fu}} and communication facilitation~\cite{Andolina2018-nf, Xia2023-pk, Liu2023-aw, Abolnejadian2025-xf}, {we have yet to encounter an integrated approach that determines \textit{when} an emerging collaborative need merits delivery and \textit{how} to synthesize retrieved content into a form amenable to synchronous meetings, where information retrieval and sensemaking are supplemental to the conversation.}

In this paper, we present \includegraphics[height=0.75em]{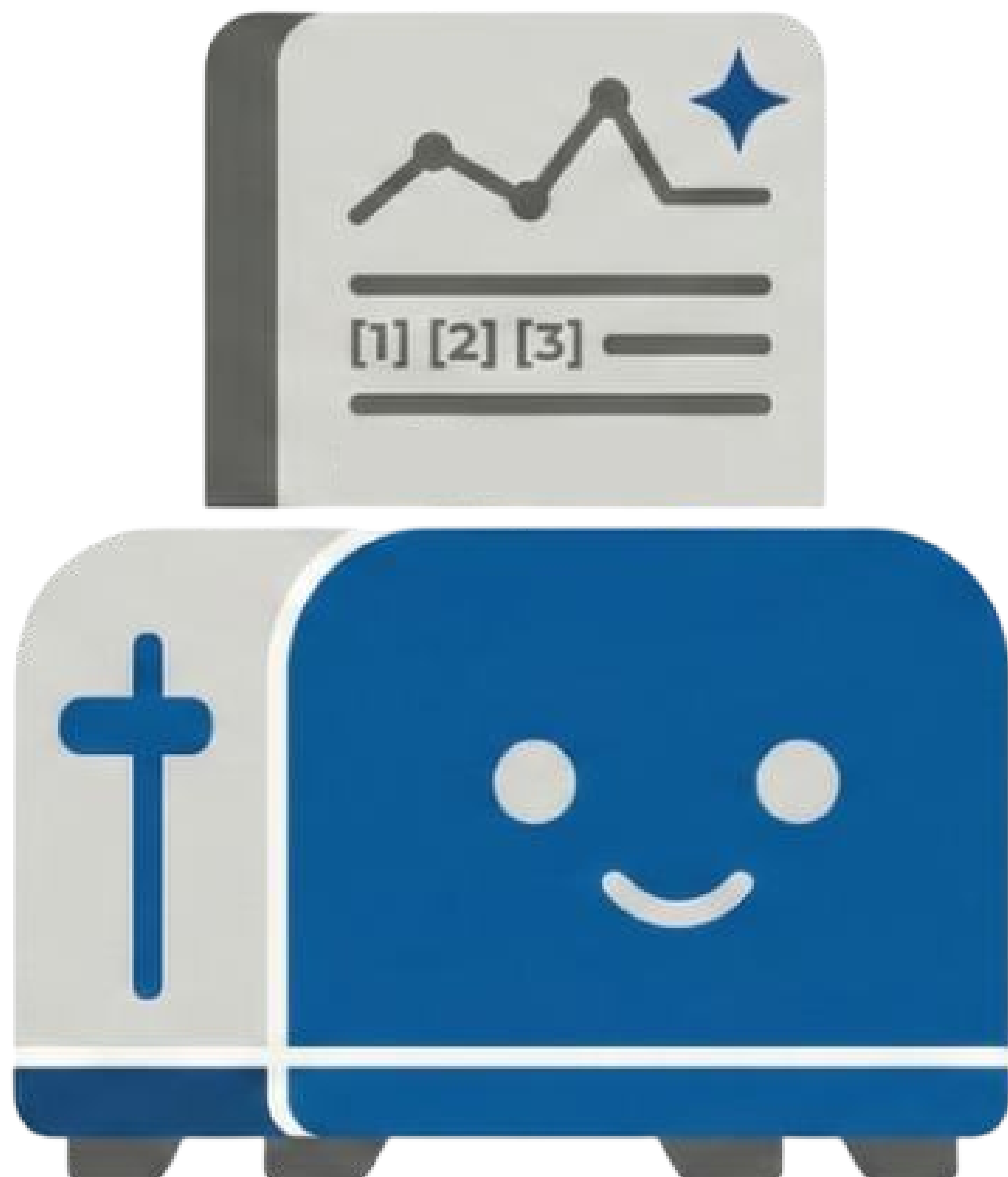} \textsc{InsightToast}, a mixed-initiative interface that monitors spoken discourse in real time, detects knowledge gaps as they emerge, and proactively delivers glanceable, source-grounded information in the form of ephemeral `toast' notifications, {which are dynamically organized in a meeting side channel according to detected meeting topics.}
{Led with a concise title that previews its content, each insight contains either a text snippet or an interactive chart prioritizing both serendipity and brevity}, traceable both to its conversational trigger and to its contributing document sources.
\textsc{InsightToast} employs a multi-agent LLM-based pipeline that tracks conversational context, detects information needs and topics as they emerge, formulates multi-faceted search queries that address complementary aspects of each identified need, and ultimately generates insights using a retrieval-augmented generation (RAG) approach that draws from an indexed knowledge base and supplementary open web sources. 
We demonstrate \textsc{InsightToast}'s application through a legislative policy deliberation scenario, collecting and processing approximately $11\,GBs$ of legislative documents from 
{Canada's House of Commons} open data and conducting a simulated meeting based on recorded committee sessions. 
We then report findings from a comparative study ($N=16$) in which participants engaged in conversations leading to their decision to support alternative legislative petitions, both using \textsc{InsightToast} and a baseline search interface. 
We found that the former's multi-faceted retrieval and proactive delivery preserved natural conversational flow, broadened the scope of deliberation, and yielded more source-grounded decision justifications, all while reducing manual retrieval effort. 
Our \textbf{contributions} include:
\begin{enumerate}
    \item The distillation of three design goals for proactive information retrieval and presentation during synchronous data-rich meetings, informed by interdisciplinary literature spanning human-computer interaction, information retrieval, visualization, and judgment and decision-making;
    \item {\textsc{InsightToast}, a mixed-initiative interface that lowers the cognitive and temporal cost of information discovery during meetings by proactively surfacing insights as glanceable, source-grounded text and interactive charts, using its multi-agent pipeline to infer \textit{what} to retrieve and \textit{when} to deliver it subject to organizational context and the pragmatics of the conversation};
    \item Findings from a mixed-methods evaluation {($N=16$)} reporting on participants' experiences with {glanceable insights surfaced proactively in the conversation's periphery during collaborative deliberation} and reflecting on the implications for integrating such capabilities into synchronous communication platforms.

\end{enumerate}



\section{Related Work}
\label{sec:related_work}

We are informed by prior work in synchronous collaboration around data, proactive information delivery, decision support under cognitive load, and AI-mediated conversational support.

\subsection{Data-Rich Synchronous Communication}

From large organizations to small collaborative settings, data and institutional knowledge serve as the basis for meetings where goals include analyzing trends \cite{Mahyar2014-vj}, brainstorming ideas \cite{Shaer2024-xc}, making decisions \cite{Dimara2022-lb}, and monitoring unfolding events \cite{Nowak2024-zu}. Foundational work in CSCW characterizes group activity along dimensions of time and space \cite{Johansen1988-qp, Ellis1991-oz}; our focus is on the synchronous side of this space.
With respect to space, our work applies to co-located, remote, and/or hybrid collaboration \cite{DeFilippis2022-pz}. 
Irrespective of where these meetings take place, and as remarked upon in the introduction, meeting participants face substantial barriers engaging with data during synchronous meetings~\cite{Brehmer2026-jo, Tory2023-gh}; live conversations routinely surface questions that expand the meeting scope beyond prepared materials~\cite{Brehmer2022-tl}, resulting in improvisation that no amount of advance curation can anticipate, the deferral of discussion to future meetings, and a risk of under-informed decisions.

\bstart{Right data, right time} 
Often the challenge is not that data is unavailable, but that it exists somewhere within an organization's reports or broader web sources \cite{Citroen2011-ti}; it is practically inaccessible in the moment it is needed. Effective synchronous collaboration depends on a continuous, low-effort stream of shared contextual information \cite{Gutwin2002-or}, a stream that is broken when participants disengage from the conversation to locate missing evidence. The result is both an individual cognitive cost, as recovering focus following interruptions can consume precious minutes \cite{Doubleday2023-bn}, as well as a collective cost, as the fluid, exploratory engagement that makes live conversations productive is brought to a halt \cite{He2021-zm}.

\bstart{Tracking the conversation} 
Keeping a coherent record of what a data-rich conversation produces is itself demanding: it requires maintaining common ground \cite{Clark2004-jp}, capturing multiple layers of analytic provenance generated through speech \cite{Ragan2016-vz}, and ensuring that partial findings can be handed off to others \cite{Zhao2018-my}. Mahyar et al. \cite{Mahyar2012-vw} observed that record-keeping in collaborative visual analytics sessions directly competed for the attentional resources needed to engage in the analysis itself. This problem extends to decision-making, where traceability between claims and their supporting data matters for accountability and decision quality \cite{Mahyar2014-vj}.

Together, these challenges shape the problem space that we address: supporting synchronous collaboration by facilitating information retrieval while keeping track of evolving topics of conversation, ideally without disrupting its flow.


\subsection{Data Retrieval and Presentation} \label{sec:related_data_presentation}

In meetings, data is typically surfaced through slide presentations, spreadsheets, and dashboards, allowing participants to reference and co-interpret data with visual aids~\cite{Brehmer2022-tl, Kim2024-hw}. While primary displays attend to these prepared data assets, prior research has examined backchannel communication as a mechanism for participants to augment understanding in parallel with the ongoing discussion \cite{Cogdill2005-vk, Dork2010-wi}. Platforms such as MeetCues \cite{aseniero2020meetcues} and CommunityClick \cite{Jasim2021-is} demonstrate how visual and interactive backchannels can improve engagement and awareness during meetings. However, engaging with backchannels imposes a dual-task cost, diverting participants' visual and cognitive attention from the primary discussion \cite{Bailey2006-af, Adamczyk2004-vi}.

\bstart{Proactive retrieval} 
A key tension arises along the dimension of information delivery within the mixed-initiative spectrum \cite{Riley1989-hj, Horvitz1999-xu}. Proactive delivery offers the potential to surface contextually relevant data without requiring people to break conversational flow \cite{Wadhwa_undated-rh, Meng2025-jo}, an idea explored by just-in-time information retrieval agents \cite{Rhodes2000-fu} and more recently by SearchBot \cite{Andolina2018-nf}, which listens to spoken conversations, detects entities, and proactively decides \textit{what} related document to surface. In the information retrieval community, frameworks \cite{Shah2018-ao} such as Information Fostering and workshops including ProActLLM \cite{Chatterjee2025-ax} have sought to formalize taxonomies for \textit{when} a system should initiate retrieval in dialogue contexts. Yet, these efforts fall short on \textit{how} to present proactively retrieved data in a manner that respects the attentional constraints of meetings.

\bstart{Glanceable and peripheral presentation} 
To mitigate this attentional cost, researchers have explored interfaces positioned at the edge of attention that convey information through ambient, low-demand encodings \cite{Weiser1995-fu, Miller2002-ix, Matthews2006-fd}. The Peripheral Displays Toolkit (PTK) \cite{Matthews2004-mh} formalizes this design space, structuring attention management around abstraction, notification levels, and display transitions. For such displays to be useful without sustained focus, they must be glanceable, conveying information during the brief peeks characteristic of constrained display contexts \cite{Blascheck2018-tc}. Within this constraint, text has been found to be among the fastest media for information extraction, followed by simple encodings leveraging pre-attentive visual features \cite{Lee2021-yi}.

Together, this body of work informs the design of \textsc{InsightToast} as a system that combines proactive, context-driven retrieval with glanceable textual and visual presentations; a pairing that, to our knowledge, has not been previously prototyped or evaluated in the context of synchronous data-driven meetings.


\subsection{Data-Driven Decision-Making} 
\label{sec:related_decision_making}
Decision-making is a cognitively demanding process that relies heavily on working memory and deliberate reasoning \cite{Sweller2011-ys, Kahneman2011-en}. In synchronous conversational settings, this load compounds, where participants must simultaneously track dialogue, process new information, and reason about trade-offs \cite{Baker2002-th, Sweller2011-ys}.
{These individual burdens extend to the group level: under-informed members tend to discuss already-shared information and withhold unshared knowledge, biasing the resulting decision \cite{Stasser1985-bp}, while reduced participation makes meetings less effective \cite{Cutler2021-nj, Hosseinkashi2024-mb}.}
These tensions motivate the need for lightweight, well-timed decision support that can operate within the constraints of an ongoing conversation \cite{Rochat2002-so}.

\bstart{Visualization for decision-makers} 
Decision-makers require concise overviews of alternatives, criteria, and trade-offs rather than exhaustive analysis \cite{Dimara2022-lb, Oral2024-py}. Despite visualization being widely positioned as a decision-support tool, prior work has found that explicit decision tasks remain underrepresented in visualization research \cite{Dimara2022-oa}, and that most tools focus on the intelligence stage (data exploration) while providing limited support for generating alternatives \cite{Oral2024-py, Dimara2018-uu}. Systems such as Dust \& Magnet \cite{Soo-Yi2005-xj} and WeightLifter \cite{Pajer2017-cy} represent targeted efforts toward multi-criteria decision support, helping people assess trade-offs interactively.

\bstart{Unknown unknowns}
A deeper challenge is not what people know they are missing, but what they do not know to look for \cite{Tversky1973-gq, Walters2023-wc}. This is compounded by the vocabulary problem \cite{Furnas1987-op}: people struggle to translate information needs into proper queries, particularly when unfamiliar with a domain \cite{Marchionini2006-iv, white2006supporting}. This dynamic also surfaces in recommendation systems, where optimizing for accuracy tends to reinforce what people already know at the expense of relevant alternatives that could reframe a decision \cite{Kotkov2023-sa, McNee2006-ur}. 

Our work draws on these threads to design lightweight, ambient data interventions that surface relevant but potentially overlooked facets of a decision problem in real time, reducing the cognitive cost of processing information while steering decision-makers toward information they would not have thought to seek.


\subsection{AI Support for Conversations}

A growing number of commercial tools now leverage AI to support conversations. Microsoft Teams \cite{MicrosoftTeamsAI2026} and Otter \cite{Otter2026} provide post-meeting summaries and action items, while platforms like Google Meet \cite{WorkspaceUnknown-rn} and Zoom's AI Companion \cite{ZoomCustomAI2026} extend this support by drawing on conversational and organizational context to answer participant questions during meetings. Tools such as Cluely \cite{Cluely2026, Bonic2025-ae} are beginning to shift toward proactive assistance by drawing on LLMs' knowledge. In parallel, research has explored how NLP and AI-driven systems can support meetings along dimensions including understanding spoken references to data \cite{Sharma2025-wm}, enabling in-meeting goal reflection \cite{Chen2025-lg}, and designing generative interfaces that bridge planning, execution, and follow-up {\cite{Park2024-jn, Vanukuru2025-iq, Park2026-yu}}.

\bstart{Mixed-initiative AI interfaces for synchronous conversation} 
Several works have explored using language models to enrich conversations with contextually relevant content. CrossTalk \cite{Xia2023-pk} creates intelligent substrates to enable dynamic annotations and in-meeting queries grounded in the live conversation. Visual Captions~\cite{Liu2023-aw} and Crosscast \cite{Xia2020-he} similarly augment spoken communication with on-the-fly visuals derived from what is being said. While these systems demonstrate the value of proactive AI support in conversational settings, they do not synthesize relevant information from a meeting-specific knowledge base into a source-traceable form, reinforcing the fragmentation between organizational knowledge and synchronous collaboration \cite{Brehmer2026-jo}.

\bstart{Retrieval-augmented generation (RAG) for conversation support} RAG \cite{Lewis2020-ds} addresses these limitations by grounding model outputs in external documents, improving factual accuracy and enabling source attribution \cite{Fokina2023-yj, Zhang2020-th}. Recent extensions handle sensemaking over larger corpora through graph-based indexing~\cite{Edge2024-ph, Gutierrez2024-np}, multi-hop reasoning~\cite{Liu2025-uo}, and agentic retrieval pipelines~\cite{Singh2025-bn}. Social-RAG \cite{Wang2024-wg} adopts this framework in asynchronous conversations by retrieving from group interaction histories to socially ground proactive suggestions. AInsight~\cite{Abolnejadian2025-xf} extends this thread to synchronous settings with a proactive, peripheral interface, though it offers limited glanceability and integration into the natural conversational flow, and lacks a formal evaluation with human subjects.

Our work builds on these foundations by designing and evaluating a multi-agent AI assistant {for during-meeting support} that actively monitors a conversation, tracks topics, identifies knowledge gaps, and delivers retrieval-augmented insights in the periphery.



\section{Design Goals}

Based on the challenges discussed in Section~\ref{sec:related_work}, we derive twelve [\textbf{D}esign \textbf{C}oncepts] organized around three [\textbf{D}esign \textbf{G}oals]: how data should be retrieved, how it should be presented, and how meeting participants maintain control and trust over surfaced information. Figure~\ref{fig:design-space} illustrates the design concepts through a motivating scenario of a co-located team meeting about market expansion.

\subsection{Proactive, Context-Aware Retrieval}
{\setlength{\fboxsep}{1pt}\colorbox{dg1bg}{\raisebox{-0.15\height}{\inlineicon{0.75\baselineskip}{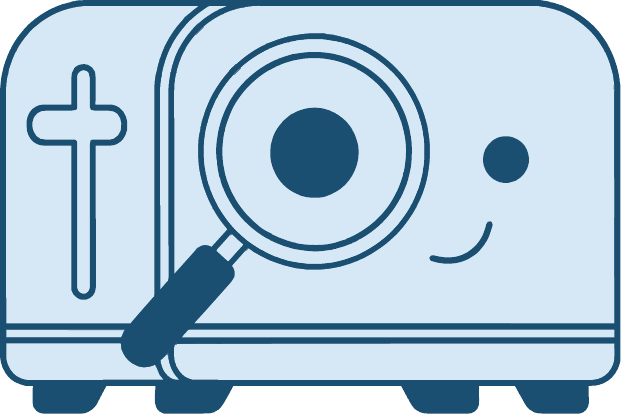}}\hspace{3pt}\textbf{\textcolor{dg1text}{DG1}}}} Manual information seeking forces participants to choose between staying engaged in discussion and pursuing the context they need \cite{Doubleday2023-bn, Brehmer2022-tl}. To avoid this trade-off, meeting support tools should autonomously monitor the discussion and trigger retrieval when participants exhibit a genuine knowledge gap, such as an explicit information request, an uncertain claim being raised, or a decision stalled for lack of evidence. Rather than issuing a single query, tools should decompose each need into multiple complementary retrievals to mitigate the vocabulary problem \cite{Furnas1987-op} and surface perspectives participants may not have considered \cite{Kahneman2011-en} (\eg~a remark about expanding to a new market might yield separate retrievals on market size, competitor presence, and regulatory landscape). Retrieval should draw from a primary organizational knowledge base supplemented by broader web sources \cite{Citroen2011-ti}, and should weight recent conversational context more heavily while remaining aware of the full discussion arc \cite{Clark2004-jp}.

\subsection{Glanceable Peripheral Presentation}  
{\setlength{\fboxsep}{1pt}\colorbox{dg2bg}{\raisebox{-0.15\height}{\inlineicon{0.75\baselineskip}{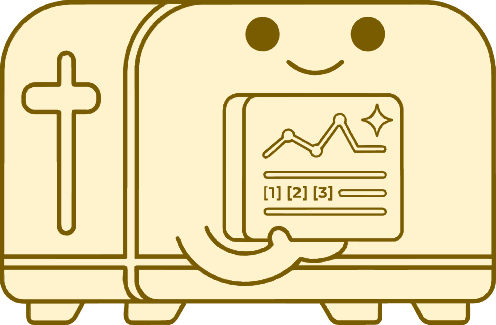}}\hspace{3pt}\textbf{\textcolor{dg2text}{DG2}}}} Content delivery must not itself become a source of disruption \cite{Bailey2006-af, Adamczyk2004-vi}. Drawing on prior work on glanceable and peripheral interfaces introduced in Section~\ref{sec:related_data_presentation}, we propose that content be delivered to a secondary display region that participants consult at their own pace, using non-modal, time-limited notifications that signal availability without demanding attention. Each piece of content should be synthesized into glanceable forms, such as concise text or compact visualizations suited to the content type, optimized for rapid comprehension during a live discussion \cite{Blascheck2018-tc, Dimara2022-lb}, and shareable across participants to maintain common ground \cite{Gutwin2002-or}. Finally, consistent with the usability principle of system status visibility \cite{nielsen1994enhancing}, retrieval and synthesis progress (initiated as in 
{\setlength{\fboxsep}{1pt}\colorbox{dg1bg}{\raisebox{-0.15\height}{\inlineicon{0.75\baselineskip}{figures/retrieval.pdf}}\hspace{3pt}\textbf{\textcolor{dg1text}{DG1}}}}) should be surfaced through ephemeral indicators, giving participants awareness of what is being generated and when to expect it, without persisting beyond its moment of relevance.

\subsection{Traceability and Control} 
{\setlength{\fboxsep}{1pt}\colorbox{dg3bg}{\raisebox{-0.15\height}{\inlineicon{0.75\baselineskip}{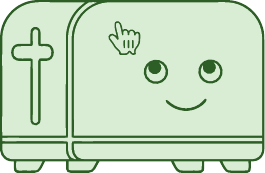}}\hspace{3pt}\textbf{\textcolor{dg3text}{DG3}}}} Surfaced content should carry dual attribution: both its retrieved data source(s) and its triggering conversational moment(s), providing accountability, grounding trust in synthesized content, and clarifying the context for judging relevance~\cite{Ragan2016-vz, Zhang2020-th}. 
The salience of information should shift with the conversation's focus, so participants should be able to curate this content, promoting useful items and dismissing irrelevant ones{, preserving human initiative over the system's proactively retrieved and synthesized results~\cite{Horvitz1999-xu}}.
Finally, the brevity of this content {\setlength{\fboxsep}{1pt}\colorbox{dg2bg}{\raisebox{-0.15\height}{\inlineicon{0.75\baselineskip}{figures/presentation.pdf}}\hspace{3pt}\textbf{\textcolor{dg2text}{DG2}}}} invites progressive disclosure (\ie~details-on-demand)~\cite{Ahlberg1994-oe}.



\section{InsightToast}
\label{sec:insight-toast}

\begin{figure*}[t]
  \centering
  \includegraphics[width=\textwidth]{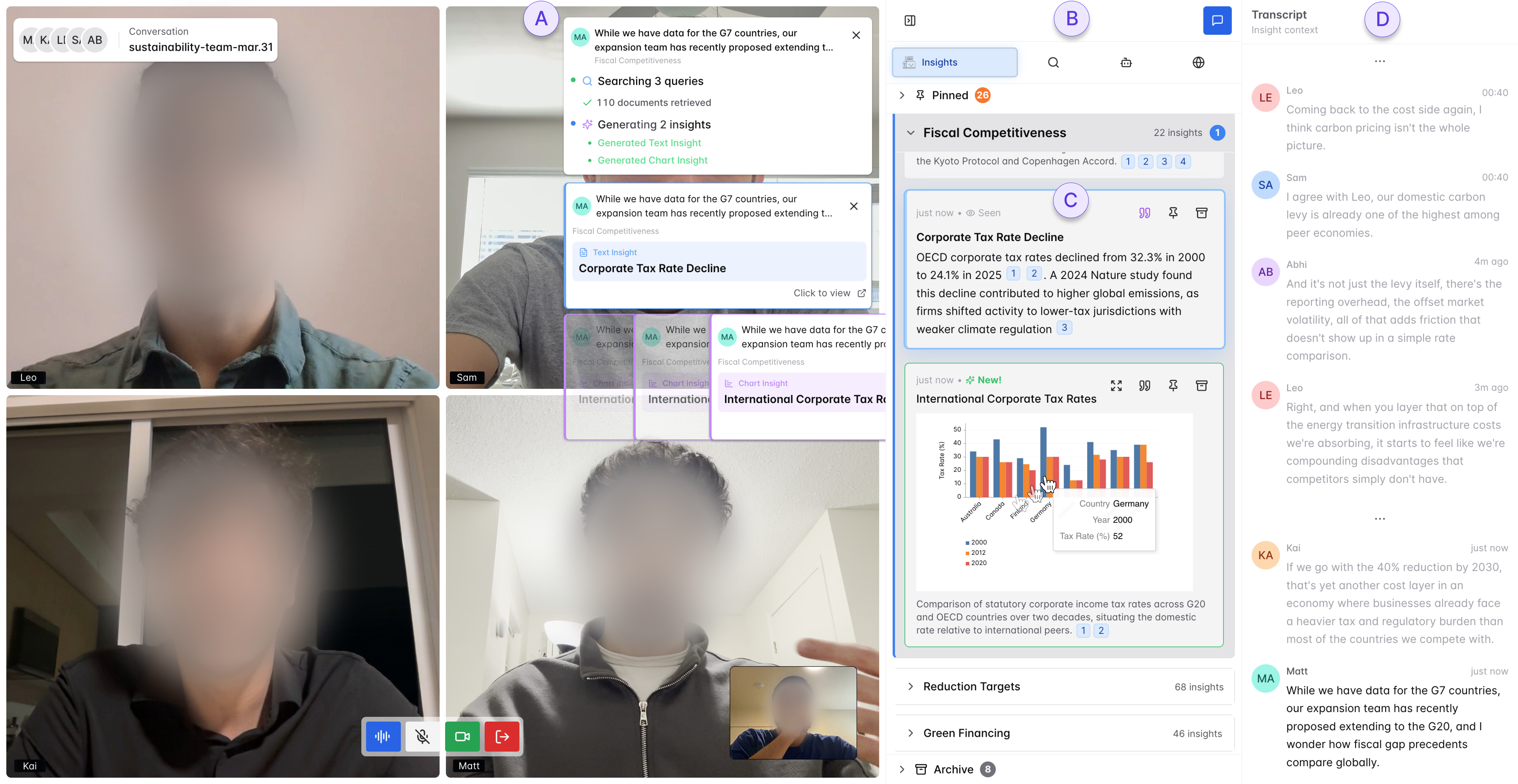}
  \caption{\textsc{InsightToast}'s interface from Abhi's view during a remote meeting. (A) Toast notifications ephemerally surface retrieval progress and synthesized insights as Matt expresses an information need, with insights organized under an expandable (B) side panel sorted into dynamically detected meeting topics. (C) Each insight appears as an interactive chart or a succinct text snippet, traceable to its retrieved document sources and originating conversational trigger via a (D) secondary transcript panel.}
  \Description{Annotated screenshot of InsightToast's interface from one participant's view during a remote meeting with five people. The left portion shows a video conferencing grid with four webcam feeds labeled Leo, Kai, Sam, and Matt, and Abhi's camera feed in the bottom right. The top-right area (labeled A) shows three stacked toast notifications reporting retrieval progress: searching three queries, 110 documents retrieved, and generating two insights, the middle toast previewing a text insight toast titled "Corporate Tax Rate Decline" and a link to view it, and the bottom toast showing a chart insight toast that is being dismissed to its right, depicted by a ghost effect. To the right (labeled B), an Insight side panel lists insights organized under dynamically detected topic headings such as "Fiscal Competitiveness" with 22 insights, "Reduction Targets" with 68, and "Green Financing" with 46, along with a Pinned section at the top and an Archive section at the bottom. The center of the side panel (labeled C) displays two insight cards: a text insight with a glowing effect on its borders representing that it was selected and a quote icon in selected mode, and below it a grouped bar chart titled "International Corporate Tax Rates" comparing rates across G20 and OECD countries for the years 2000, 2012, and 2025, with a tooltip showing exact values for Germany. A secondary Transcript panel (labeled D) on the far right shows a running conversation log with speaker-attributed utterances and timestamps, ending with Matt's remark that triggered the retrieval and is displayed in a highlighted font color because the quote icon was selected.}
  \label{fig:ui}
\end{figure*}

Guided by the aforementioned design goals, we designed and developed \inlineicon{0.75em}{figures/logo.pdf} \textsc{InsightToast}, a system that actively monitors live conversation, tracks topics as they emerge and evolve, and detects information gaps to proactively generate representations of this information delivered to the conversation's side channel; as a goal of these deliveries is serendipity, we hereafter refer to them as \textit{insights}. Each insight is a succinct, synthesized information snippet grounded in the meeting's knowledge base and supplemented by the open web, traceable to both its conversational trigger and its underlying data sources and affording curation control over them.

To anchor our explanation, we follow a running example throughout this section in which a sustainability team at an energy company deliberates on an emission-reduction target, with the meeting's knowledge base comprising internal organizational documents and a corpus of national legislative and regulatory records.

\subsection{Interface and Interaction}
\label{sec:interface}

To realize {\setlength{\fboxsep}{1pt}\colorbox{dg2bg}{\raisebox{-0.15\height}{\inlineicon{0.75\baselineskip}{figures/presentation.pdf}}\hspace{3pt}\textbf{\textcolor{dg2text}{DG2}}}} and {\setlength{\fboxsep}{1pt}\colorbox{dg3bg}{\raisebox{-0.15\height}{\inlineicon{0.75\baselineskip}{figures/control.pdf}}\hspace{3pt}\textbf{\textcolor{dg3text}{DG3}}}}, \textsc{InsightToast}'s interface juxtaposes two side panels with a main panel; the latter is either a video-conferencing view for remote collaboration (Figure~\ref{fig:ui}), or a participant's desktop environment in co-located scenarios (Figure~\ref{fig:teaser}).
Inspired by macOS's notification center placement \cite{apple_notification_center}, ephemeral \emph{Insight Toasts} appear as notification overlays in the top-right corner of the conversation's main panel, conveying both real-time progress on information retrieval and a concise title for each newly synthesized insight. An \emph{Insight Sidebar} provides an expandable panel that aggregates all generated insights, organized under dynamically identified meeting topics, allowing meeting participants to explore and curate synthesized content at their own pace.

\bstart{Toasts}
Given \textsc{InsightToast}'s system-initiated retrieval, participants must remain aware of when new content becomes available. We employ \emph{toasts} as transient, non-modal, time-based UI components \cite{carbon2023notification} to communicate system status \designconcept{figures/control.pdf}{DC2.2}{dg2bg}{dg2text}. \emph{Progress Toasts} appear once the system detects a knowledge gap and updates incrementally throughout retrieval and synthesis, surfacing high-level status indicators such as the number of source documents retrieved and considered for synthesis \designconcept{figures/control.pdf}{DC2.4}{dg2bg}{dg2text}. 
\emph{Insight Toasts} appear upon the completion of synthesis {as the first level of insight disclosure}, displaying a title of up to five words {that previews the synthesized content while limiting conversational disruption \cite{Ofek2013-ok}} alongside speaker attribution and its conversational moment. Participants can click the toast to be directed to the insight's full representation in an Insight side panel (Figure~\ref{fig:ui}.A).
Depending on the structure of the retrieved data, the system's pipeline selects the appropriate presentation modality (see Section~\ref{sec:pipeline}), transforming evidence into glanceable, source-grounded textual or visual insights.

\emph{Text Insights} summarize relevant unstructured evidence. They lead with the most salient finding, anchor claims in temporal and institutional context (\eg~specifying when testimony was given or which board meeting a report originates from), and attach inline citations so that every claim can be traced to its underlying document \designconcept{figures/control.pdf}{DC3.1}{dg3bg}{dg3text}. To support glanceability, each text insight is constrained to a tweet-length snippet capped at 280 characters, a length concise enough to convey a self-contained message, yet short enough to be peeked at under five seconds \cite{Counts2021-zk}, consistent with the glance durations observed for constrained peripheral displays \cite{Lee2021-yi} \designconcept{figures/presentation.pdf}{DC2.3}{dg2bg}{dg2text}.
Meanwhile, \emph{Chart Insights} encode relevant structured evidence in a compact form while embedding finer detail within an interactive layer accessible on demand \designconcept{figures/control.pdf}{DC3.4}{dg3bg}{dg3text}. Informed by the glanceable visualization literature, \emph{Chart Insights} are drawn from a restricted chart inventory favoring encodings known to support rapid human judgment: length along a common scale for quantities (bar charts) and temporal patterns (line charts), and arc length for part-to-whole relationships (donut charts)~\cite{Blascheck2018-tc}. To preserve legibility within the compact peripheral display, the system enforces high data-ink ratios \cite{Inbar2007-lj} by suppressing non-essential elements such as gridlines, redundant legends, and auxiliary annotations. 

In Figure~\ref{fig:ui}, which captures a moment in our sustainability meeting scenario, Kai raises a potential tax concern, after which Matt poses an information need beyond the data readily available in the conversation, stating \textit{``I wonder how fiscal gap precedents compare globally.''} Drawing on both meeting and conversational moment context, \textsc{InsightToast} surfaces serendipitous insights comparing corporate tax rate trends at a global scale, notifying all participants of retrieval and synthesis progress through dismissable toast notifications (Figure~\ref{fig:ui}.A). Abhi, choosing to explore the synthesized insights further, clicks on an insight toast, which expands the Insight Sidebar and brings the full content into his personal view. There, he can compare tax rate trends among major economies at a glance, with the ability to inspect exact values by hovering over any data point in the chart (Figure~\ref{fig:ui}.C).

\begin{figure*}[t]
  \centering
  \includegraphics[width=\textwidth]{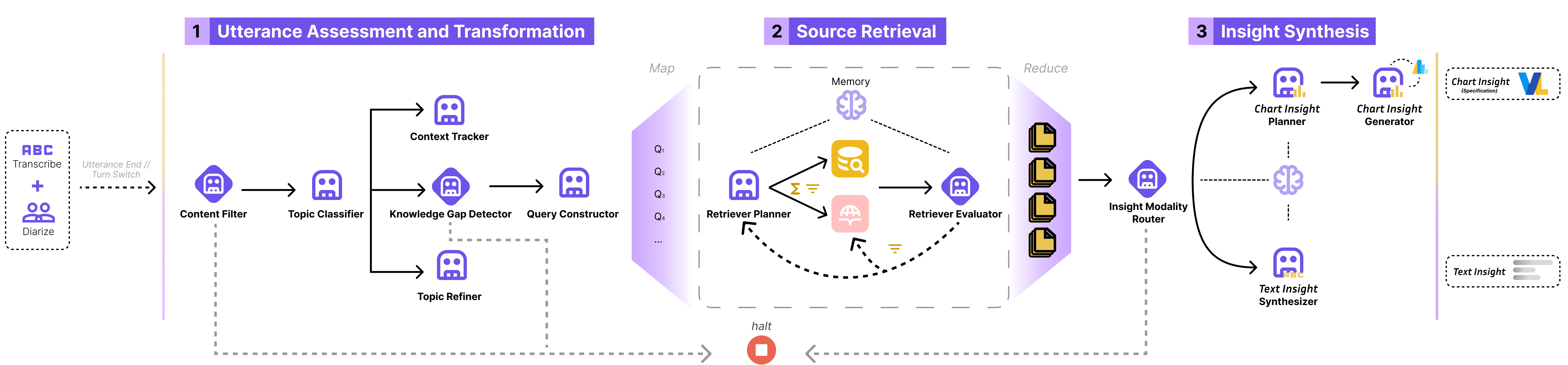}
  \caption{\textsc{InsightToast}'s multi-agent LLM-based pipeline processes diarized speech utterances sequentially through a three-stage process: (1) maintaining a running conversational context to facilitate topic tracking, (2) multi-faceted proactive retrieval across indexed and web sources, and (3) insight synthesis.}
  \Description{Architecture diagram of InsightToast's multi-agent retrieval and synthesis pipeline, flowing left to right through three labeled stages. The input on the far left is a transcribed and diarized speech utterance, which triggers the pipeline on either the utterance end or the turn switch between speakers. Stage 1, Utterance Assessment and Transformation, begins with a Content Filter routing agent, which either halts the pipeline or passes to a Topic Classifier. The Topic Classifier is connected to three parallel Context Tracker, Knowledge Gap Detector, and Topic Refiner agents. The Knowledge Gap Detector agent acts as the primary gate to proactive retrieval initiation and either halts the pipeline or moves the pipeline to the Query Constructor agent for utterances that are recognized to need an insight, leading to multiple search queries being spawned considering the current context of the conversation. Stage 2, Source Retrieval, follows a map-reduce pattern: multiple queries are dispatched in parallel to a Retriever Planner agent [Map] that accesses both an indexed knowledge base and web sources, followed by a Retriever Evaluator agent that can loop back to the planner for iterative refinement or halt the loop if information is sufficient and send all of the documents retrieved from parallel subgraphs to the next stage [Reduce]. Stage 3, Insight Synthesis, starts with an Insight Modality Router agent that either halts the pipeline if retrieved documents are insufficient or irrelevant, or decides on generating either a chart insight, a text insight, or both. A Chart Insight Planner followed by a Chart Insight Generator producing a chart output, and a Text Insight Synthesizer producing a text output follow. All agents share a stateful memory.}
  \label{fig:pipeline}
\end{figure*}

\bstart{Side panels}
Each participant can access a side panel that can be expanded when further context is needed \designconcept{figures/control.pdf}{DC2.1}{dg2bg}{dg2text}, either by browsing synthesized insights under their corresponding topics or by performing reactive retrieval through a direct knowledge-base search, an LLM chatbot, or a web search engine, complementing \textsc{InsightToast}'s proactive retrieval. The panel helps participants maintain awareness of subjects discussed during a meeting via topic tracking \designconcept{figures/control.pdf}{DC3.2}{dg3bg}{dg3text}. Without requiring meeting participants to define topics at the outset, the system automatically identifies and creates distinct topic categories as the conversation unfolds, transitioning between them as discussion shifts and surfacing the active topic's content without requiring manual navigation. To afford control over proactively generated content, participants can collaboratively curate insights by archiving those deemed irrelevant or pinning those particularly valuable to the deliberation into shared pin and archive collections \designconcept{figures/control.pdf}{DC3.3}{dg3bg}{dg3text}. To facilitate collaboration around a specific piece of synthesized information, any participant can bring an insight into focus across all participants' views by selecting it \designconcept{figures/control.pdf}{DC2.5}{dg2bg}{dg2text} (Figure~\ref{fig:ui}.C). The primary insight side panel is complemented by a secondary transcript side panel, which allows meeting participants to reference specific conversational moments at three levels of detail: \textit{(1)} the full running transcript, \textit{(2)} utterances belonging to a specific topic by expanding that topic in the insight side panel, or \textit{(3)} the specific conversational trigger of an insight \designconcept{figures/control.pdf}{DC3.1}{dg3bg}{dg3text}.

\subsection{Proactive Retrieval and Synthesis Pipeline}
\label{sec:pipeline}

\textsc{InsightToast} employs a three-stage natural language processing pipeline (Figure~\ref{fig:pipeline}) that continuously ingests transcribed and diarized speech utterances.
It determines when, what, and how to retrieve and synthesize relevant information, collectively answering the requirements posed by 
{\setlength{\fboxsep}{1pt}\colorbox{dg1bg}{\raisebox{-0.15\height}{\inlineicon{0.75\baselineskip}{figures/retrieval.pdf}}\hspace{3pt}\textbf{\textcolor{dg1text}{DG1}}}}. As the pipeline's objective is to support a running meeting from start to finish, it processes input incrementally, maintaining an evolving picture of the broader discussion so that later turns can still be interpreted in light of earlier references and topics. The pipeline is realized as a multi-agent LLM-based directed graph in which specialized agents operate over a shared state, enabling task specialization, independent verification of intermediate results, and distributed reasoning 
that collectively surpass the capacity of a single monolithic language model \cite{Guo2024-fa, Tran2025-pp, Du2023-hs}, a pattern demonstrated across different frameworks \cite{Wu2023-pa} and application domains \cite{Hong2023-wu}. 

\bstart{(i) Utterance assessment and transformation}
Upon receiving a transcribed utterance, a content filter agent, informed by the DAMSL dialogue act annotation scheme \cite{Core1997-ue}, separates communication management utterances from task-level content.
Retained segments are either assigned to an existing topic or instantiate a new one via few-shot classification \cite{Brown2020-nr, Sun2023-xo},
while a meeting-level tracker maintains a running summary and key factual anchors (e.g., stated targets, participant roles, decisions in progress) to keep all downstream agents context-aware across assessment, retrieval, and synthesis. Using this context built around the utterance, the pipeline next evaluates whether the current moment exhibits a genuine knowledge gap \designconcept{figures/retrieval.pdf}{DC1.1}{dg1bg}{dg1text} {by weighing the expected utility of surfacing new information against the cost of an unsolicited toast \cite{Horvitz1999-xu}, operationalized through five trigger categories (detailed in Appendix~\ref{app:gap-criteria})}, drawing on established information-seeking signals such as Belkin's ASK model \cite{Belkin1980-bm}, 
linguistic hedges as markers of epistemic uncertainty \cite{Lakoff1973-pk}, and clarification requests signaling incomplete understanding \cite{Purver2003-gh}. 
When a gap is confirmed, a query construction agent decomposes the underlying information need into multiple complementary search queries \cite{Zhou2022-tn}, each augmented with hypothetical document descriptions \cite{Gao2022-eu} to improve semantic alignment while surfacing perspectives participants may not have considered \designconcept{figures/retrieval.pdf}{DC1.2}{dg1bg}{dg1text}. 
For instance, a remark such as \textit{``I'm not sure what regulatory risks would come with a more aggressive target?''} is decomposed into queries targeting \textit{federal and sub-national reporting requirements}, \textit{feasibility constraints associated with accelerated emissions reductions}, and \textit{prior ambitious enforcement precedents}, three aspects the team may not have had the time or thought to seek individually, before being dispatched to the retrieval stage.

\bstart{(ii) Source retrieval} 
Queries are dispatched to parallel subgraphs following a map\textendash reduce pattern \cite{Dayalan2008-je}. Within each subgraph, a planner agent refines the query with the help of synthetic hypothetical documents, applies metadata filters, calibrates retrieval scope, and determines whether to route retrieval to the meeting's indexed knowledge base, a web search API, or both \designconcept{figures/retrieval.pdf}{DC1.3}{dg1bg}{dg1text}. Following an adaptive retrieval strategy \cite{Jeong2024-zn, Yan2025-ed}, a secondary feedback agent evaluates the retrieved document set for sufficiency and relevance, adjusting the query and re-executing retrieval in an iterative refinement loop \cite{Joren2024-br} that terminates once adequate coverage is confirmed or a fixed iteration ceiling is reached.
Once all subgraphs conclude, their results are merged into a unified, ranked evidence set.

\bstart{(iii) Insight synthesis} A routing agent assesses the aggregated evidence against the current conversational context to determine whether to surface a new insight, groups queries addressing facets of the same information need, and routes each group to the appropriate output modality: a textual summary, a chart, or both.
\emph{Text Insights} are generated following a RAG approach \cite{Gao2023-up}, ensuring that every claim remains traceable to its underlying source document. \emph{Chart Insights} follow a two-phase plan-then-generate process inspired by stepwise chart reasoning \cite{Tian2023-ys}, in which a planning agent first determines the primary analytic intent of the evidence, either comparison, trend, composition, or ranking, and selects a chart type from a restricted inventory to support glanceability{;  Appendix~\ref{app:chart-constraints} provides more detail about this process}.
A generation agent then produces the chart as a declarative Vega-Lite \cite{Satyanarayan2017-ee} specification by writing and executing Altair \cite{VanderPlas2018-ok} code in a sandboxed environment, leveraging interleaved reasoning and action \cite{Yao2022-vl} to iteratively verify both the structural validity of the specification and the semantic fidelity of the rendered output.

\subsection{Implementation}
\label{sec:implementation}

\textit{InsightToast}'s backend implements the pipeline described in Section~\ref{sec:pipeline}, exposed via a RESTful API to the front-end interface detailed in Section~\ref{sec:interface}. The backend is built on LangGraph \cite{langgraph2024} as the agent orchestration framework, with Gemini Flash Lite Latest (2.5 Flash Lite at development time) serving as the language model for the majority of pipeline agents and Gemini Flash Latest (3 Flash at development time) assigned to the chart planner and generation agents given its stronger tool-use capabilities. The API layer is implemented in FastAPI \cite{fastapi2024}, providing an endpoint that accepts transcribed speech utterances and streams asynchronous updates to connected clients. The front-end is a React web application using Chakra UI \cite{chakraui2024} as its component library and LiveKit \cite{livekit2025} for video conferencing. For co-located settings, the interface is embedded within a simulated macOS desktop environment \cite{quanla_macosdemo} to emulate its integration within participants' native environments. Automatic Speech Recognition (ASR) is handled by Speechmatics \cite{speechmatics2024}, which provides speaker identification alongside transcription to fulfill the pipeline's diarized utterance requirement. {The source code is available at \href{https://github.com/ubixgroup/InsightToast}{\texttt{github.com/ubixgroup/InsightToast}}.}



\section{Evaluation}
\label{sec:evaluation}

To empirically evaluate our approach, we conducted a within-subjects comparative study. 
In a {\raisebox{-0.09em}{\inlineicon{0.8em}{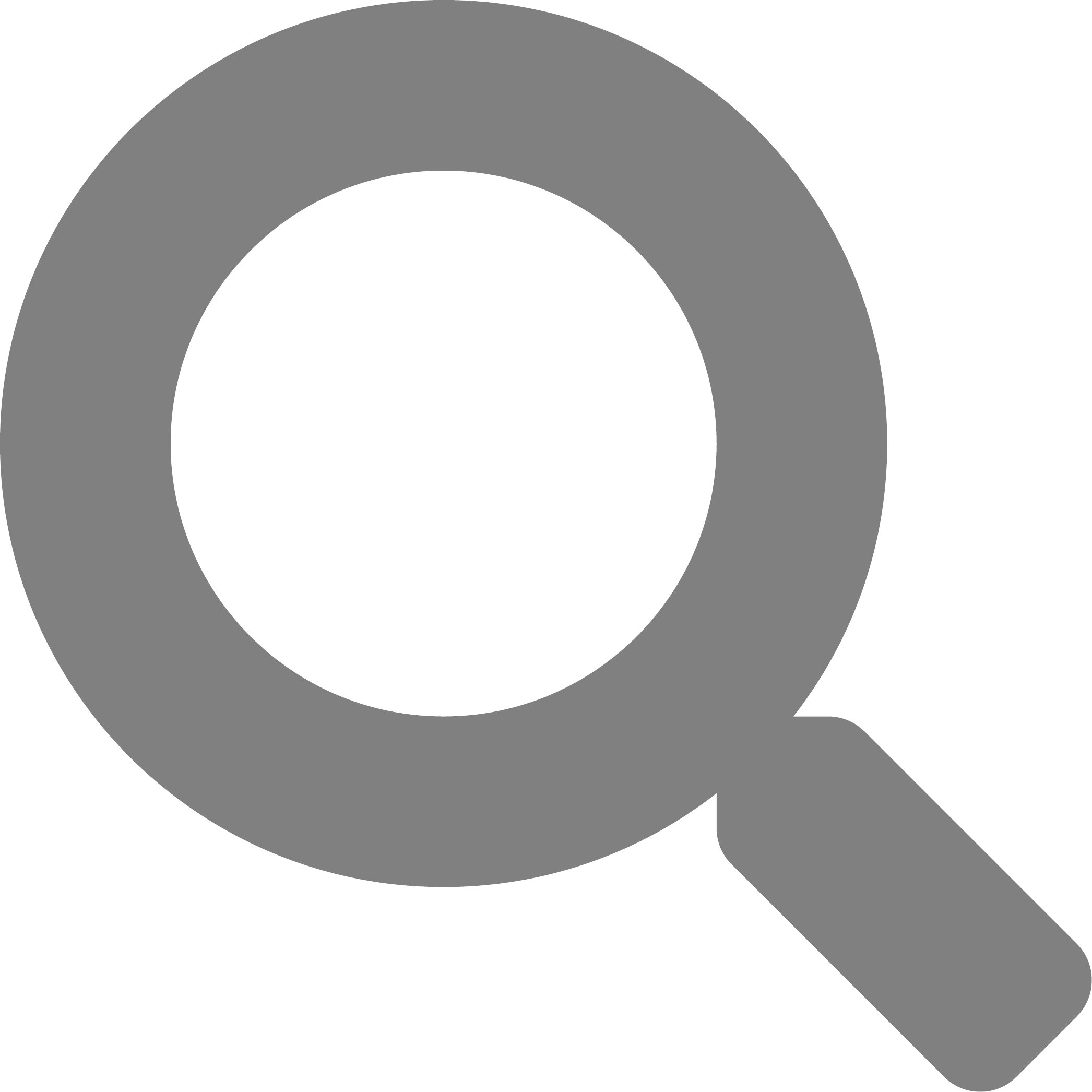}} \textit{\underline{B}ase\underline{L}ine} (\textit{BL})}{} condition, we provided participants with an interface that embedded three reactive search widgets consisting of direct knowledge base search, a chatbot (powered by the Gemini API), and web search in the conversation's side channel to approximate current practices for retrieving information during meetings.
In the {\inlineicon{0.75em}{figures/logo.pdf} \textit{\underline{I}nsight\underline{T}oast} (\textit{IT})}{} condition, we provided participants with the interface described in Section~\ref{sec:interface}, to which we added the same reactive search capabilities available in the {\raisebox{-0.09em}{\inlineicon{0.8em}{figures/magnifier.pdf}} \textit{\underline{B}ase\underline{L}ine} (\textit{BL})}{} interface. 

\begin{figure}[h!]
	\centering
	\includegraphics[width=0.47\textwidth]{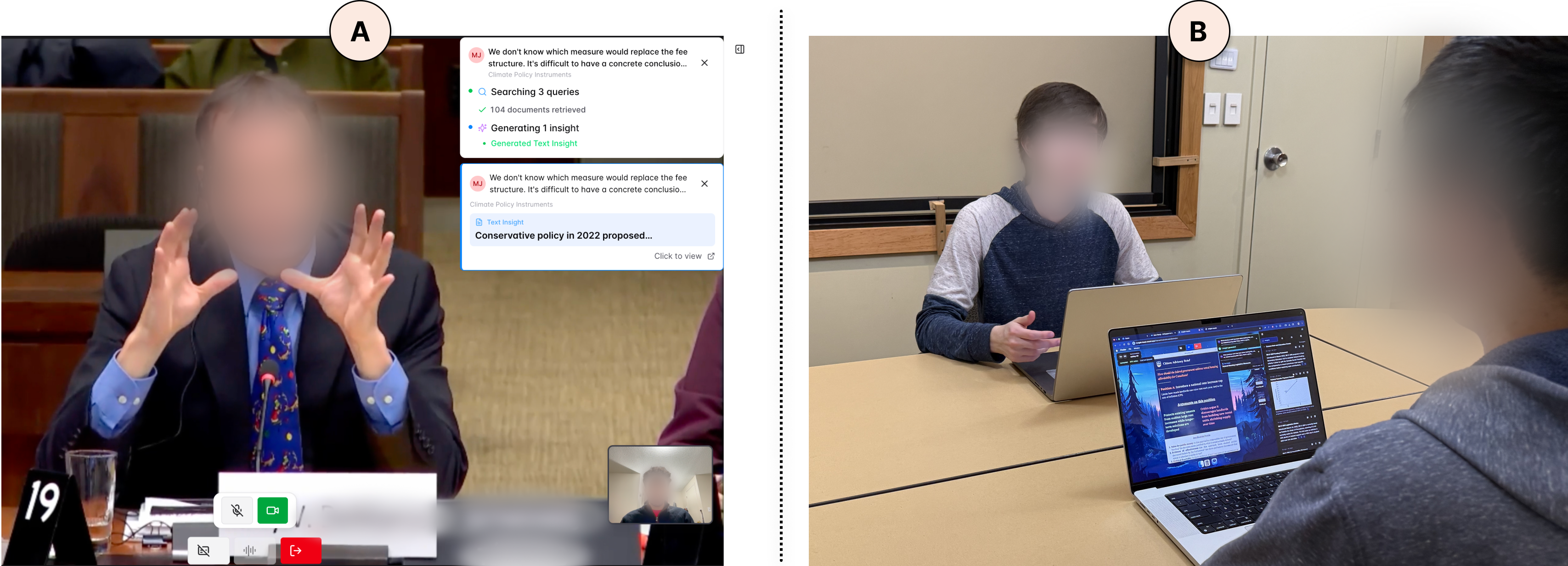}
	\caption{(A) We first demonstrated \textsc{InsightToast}'s functionality to study participants by synchronizing it with the playback of an open government webcast recording. The main study protocol accommodated both remote (Figure~\ref{fig:ui}) and  (B) face-to-face conversations.}
    \Description{(A) Screenshot of InsightToast synchronized with a government webcast recording during the study demo phase. A video feed shows a committee member speaking with a gesturing hand, while toast notifications overlay the top-right corner, reporting retrieval progress of searching three queries, 104 documents retrieved, and generating one text insight titled "Conservative policy in 2022 proposed." 
    (B) Photo of the co-located study setup showing two people seated at a table in a room (Actors). A participant faces a laptop displaying the InsightToast interface for co-located meetings, sitting across from a member of the research team who serves as the conversation partner.}
	\label{fig:task-setting} 
\end{figure}

\bstart{Scenario and knowledge base}
Policy deliberations typically draw heavily on precedents and accumulated evidence to inform decision-making. 
Legislative committee meetings exemplify this pattern, as they are accompanied by extensive context, including prior reports, research briefings, and legislative documents such as bills and amendments. These sessions are routinely published alongside their supporting materials as part of governmental transparency mandates, providing both recorded deliberations and a rich, openly available knowledge base well-suited for developing and evaluating \textsc{InsightToast} on real-world, data-rich meetings. 
Through web scraping and open APIs \cite{openparliament, houseofcommons_canada, libraryofparliament}, we collected over 40 thousand records spanning Hansard debates, petitions, recorded votes, and related legislative material from {Canada's parliamentary open data sources} and indexed them into a Qdrant vector database \cite{qdrant2024}. 
We then used this indexed knowledge base to simulate \textsc{InsightToast} across multiple recorded committee sessions, informing development and serving as a live demo during our studies. {The collected dataset, together with its processed and embedded vector store used as the knowledge base in this evaluation is open sourced at} \href{https://zenodo.org/records/21502545}{\texttt{zenodo.org/records/21502545}}.


\bstart{Task}
In each task, we asked participants to assume the role of a citizen advisor{, a lay stakeholder recommending a policy position to inform a legislative decision,} to support one of two competing federal legislative petitions {on a healthcare or housing policy issue} through deliberation with a conversation partner (a member of the research team).
{We assigned one such task per condition to mitigate learning effects, counterbalancing both the task-condition pairing and presentation order with a Latin square yielding four orderings assigned evenly across participants.}
{After the deliberation,} we asked participants to write a justification for their decision based on financial validity, precedents of effectiveness, and political feasibility.
{This post-conversation justification served to externalize participants' decision process, requiring them to deliberately ground their recommendation in the meeting's knowledge base.}
We provide the task briefs in Appendix~\ref{app:study-task}.

\bstart{Procedure}
Each study session lasted approximately 80 minutes, {beginning by obtaining informed consent from all participants}.
We \textbf{(1)} first asked participants to comment on their current practices for collaborative decision-making, information retrieval, and discourse tracking during meetings (5 min). 
\textbf{(2)} Next, we provided a guided walkthrough of the system and introduced the interface through a recording of a recent federal committee meeting on environment and sustainable development as depicted in Figure~\ref{fig:task-setting}.A (retrieved from the government's official webcasting repository).
This allowed participants to observe \textsc{InsightToast} as a passive meeting attendee in a realistic data-rich meeting context (20 min). 
\textbf{(3)} Following the demo, participants completed both study conditions: reviewing the task cards, engaging in a 15-minute deliberation conversation, making their decision, writing their justification, and completing a post-task questionnaire. 
The conversation partner used scripted open-ended cues
(\eg~\textit{``Has any bill on national rent capping made it through the legislative assembly?''}{; a list of cues appears in Appendix~\ref{app:task-script}})
to build rapport and to ensure that the conversation did not veer off-topic. 
The questionnaire probed the perceived information retrieval and decision-making effectiveness, integrating questions from information retrieval~\cite{Kelly2009-nz} and Satisfaction with Decision (SWD)~\cite{Holmes-Rovner1996-xy} evaluation as well as workload~\cite{Hart1988-px}, conversational engagement, and system usability.
\textbf{(4)} The session concluded with a semi-structured interview in which participants compared their experience across conditions and reflected on the viability of \textsc{InsightToast} for their existing meeting practices (15 min).

We deployed the {\inlineicon{0.75em}{figures/logo.pdf} \textit{\textit{IT}}}{} and {\raisebox{-0.09em}{\inlineicon{0.8em}{figures/magnifier.pdf}} \textit{BL}}{} interfaces on Vercel, communicating with a backend hosted on a server where all session logs were collected. 
The study was approved by {University of Waterloo}'s research ethics board {(approval number \#47676)}, and participants were remunerated with a {\$25 CAD} multi-retailer gift card.

\bstart{Participants}
We recruited 16 participants (8 men, 8 women; age: $M = 24.4$, $SD = 3.4$, range 19--35; all fluent English speakers) through purposive convenience sampling via departmental mailing lists.
Participants were predominantly graduate students (13 graduate, 3 undergraduate), with backgrounds spanning computer science ($n=11$; including HCI, information visualization, usable security, and data science) and policy and management sciences ($n=5$; including public service, political science, and management science engineering). 
To ensure the sample reflected a range of prior exposure to the study's core activities, we stratified recruitment along two dimensions: familiarity with federal politics (2 very familiar, 7 somewhat, 5 slightly, 2 not at all) and frequency of participation in meetings involving data (7 several times per week, 8 a few times per month, 1 rarely). 
Eleven participants attended in person (Figure~\ref{fig:task-setting}.B showing the co-located setup), while five joined remotely using screens ranging from 13--14\,in ($n=2$), 15–26\,in ($n=2$), to 27\,in or larger ($n=1$), providing variation to account for potential effects of screen size on the interface experience. 

\begin{figure*}[h!]
  \centering
  \includegraphics[width=\textwidth]{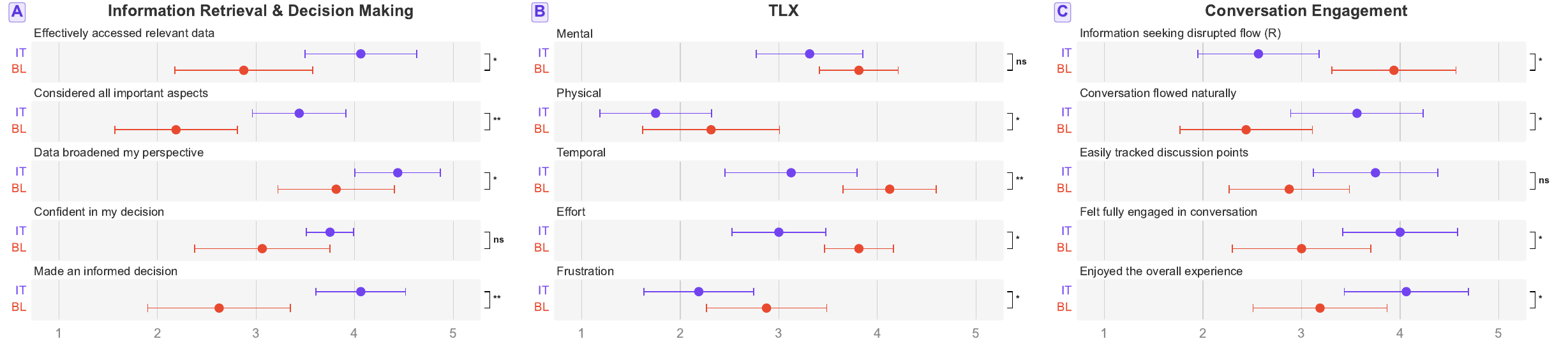}
  \caption{Questionnaires in our study asked participants to compare {\inlineicon{0.75em}{figures/logo.pdf} \textit{\underline{I}nsight\underline{T}oast} (\textit{IT})}{} and {\raisebox{-0.09em}{\inlineicon{0.8em}{figures/magnifier.pdf}} \textit{\underline{B}ase\underline{L}ine} (\textit{BL})}{} interfaces with respect to information retrieval~\cite{Kelly2009-nz}, decision satisfaction~\cite{Holmes-Rovner1996-xy}, workload~\cite{Hart1988-px}, engagement, and usability.}
  \Description{(A) Dot plots comparing InsightToast (IT) and BaseLine (BL) on five Information Retrieval and Decision Making measures on a 1 to 5 Likert scale. IT scores higher than BL on all five measures: effectively accessed relevant data (IT approximately 4.0, BL approximately 2.8, significant), considered all important aspects (IT approximately 3.4, BL approximately 2.2, highly significant), data broadened my perspective (IT approximately 4.4, BL approximately 3.8, significant), confident in my decision (IT approximately 3.8, BL approximately 3.4, not significant), and made an informed decision (IT approximately 4.1, BL approximately 2.6, highly significant). 
  (B) Dot plots comparing IT and BL on five NASA Task Load Index (TLX) dimensions on a 1 to 5 scale, where lower scores indicate less workload. Mental demand shows no significant difference with similar means around 3.5. IT scores significantly lower than BL on physical, temporal (IT approximately 3.1, BL approximately 4.1, highly significant), effort, and frustration. 
  (C) Dot plots comparing IT and BL on five Conversation Engagement measures on a 1 to 5 Likert scale. The first item, "information seeking disrupted flow," is reverse-coded, where IT scores significantly lower than BL. IT scored significantly higher on conversation flowed naturally, felt fully engaged, and enjoyed the overall experience. Easily tracked discussion points show no significant difference. IT's mean results generally range from 3.5 to 4.2 while BL's mean results range from 2.5 to 3.2 (on non-reverse coded results).}
  \label{fig:results}
\end{figure*}

\subsection{Findings}

\begin{figure}[h!]
	\centering
	\includegraphics[width=0.47\textwidth]{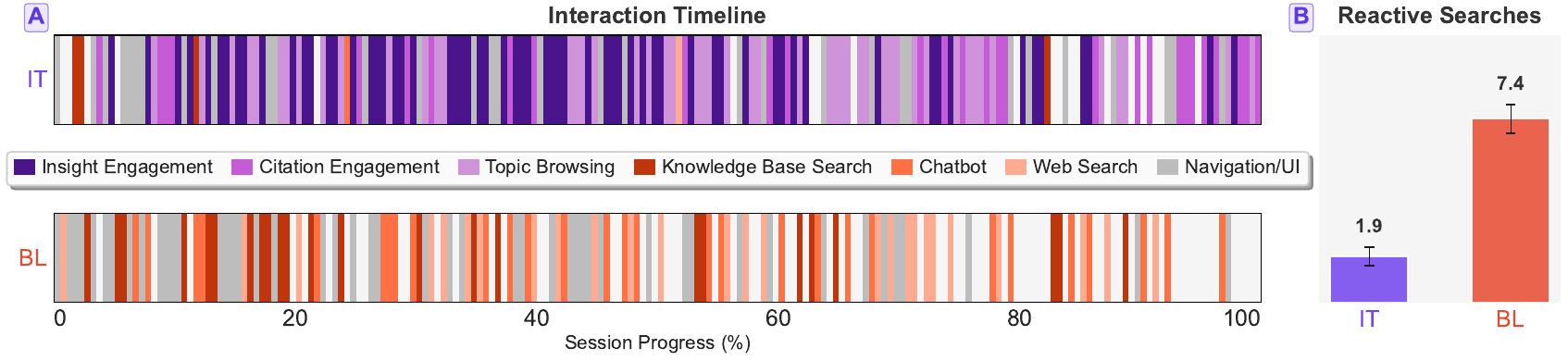}
	\caption{(A) Aggregated interaction frequencies and types are compared between the {\inlineicon{0.75em}{figures/logo.pdf} \textit{\textit{IT}}}{} and {\raisebox{-0.09em}{\inlineicon{0.8em}{figures/magnifier.pdf}} \textit{BL}}{} interfaces across all study sessions. (B) Overall, participants performed far fewer reactive searches (knowledge base search, chatbot interaction, and Google search) with {\inlineicon{0.75em}{figures/logo.pdf} \textit{\textit{IT}}}{} relative to with {\raisebox{-0.09em}{\inlineicon{0.8em}{figures/magnifier.pdf}} \textit{BL}}{}.}
    \Description{(A) Interaction timeline showing aggregated interaction events as color-coded vertical bars across session progress from 0 to 100 percent, for both InsightToast (IT) and BaseLine (BL) conditions. Seven interaction types are color-coded: Insight Engagement, Citation Engagement, Topic Browsing, Knowledge Base Search, Chatbot, Web Search, and Navigation/UI. The IT timeline is dominated by purple hues throughout, indicating that most interactions involved engaging with proactively surfaced insights, citations, and topics. The BL timeline is dominated by red-orange hues in the first 60 percent, reflecting frequent reactive searches, followed by sparse gray segments and visible gaps with no interaction in the latter 40 percent.
    (B) Bar chart comparing mean number of reactive searches per session between InsightToast (IT) and BaseLine (BL) conditions. IT averages 1.9 reactive searches while BL averages 7.4, nearly four times higher.}
	\label{fig:exp1-task} 
\end{figure}

\begin{figure}[h!]
	\centering
	\includegraphics[width=0.47\textwidth]{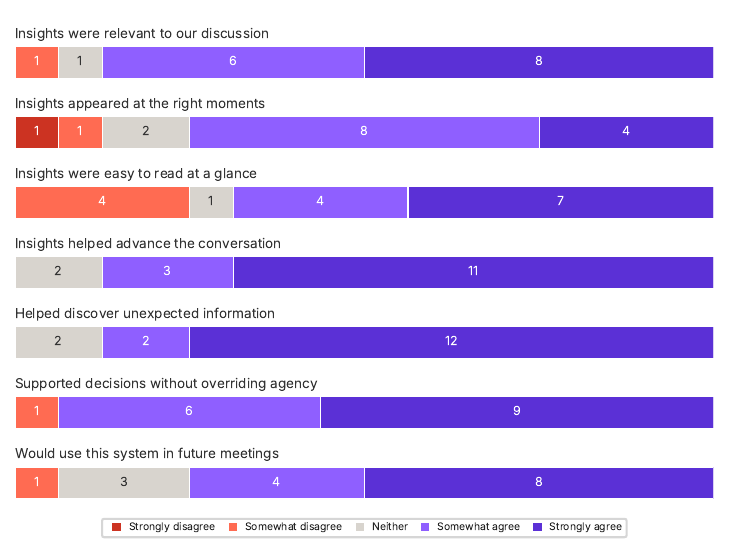}
	\caption{{Participants' agreement with statements regarding the perceived utility of \textsc{InsightToast} ($N{=}16$).}}
    \Description{Horizontal stacked bar chart showing participants' responses (N=16) on a 5-point Likert scale, color-coded from dark red (strongly disagree) through salmon (somewhat disagree), grey (neither), light purple (somewhat agree), to dark purple (strongly agree), across seven statements about InsightToast. The following lists each statement and the number of participants selecting each response option: "Insights were relevant to our discussion": 1 somewhat disagree, 1 neither, 6 somewhat agree, 8 strongly agree. "Insights appeared at the right moments": 1 strongly disagree, 1 somewhat disagree, 2 neither, 8 somewhat agree, 4 strongly agree. "Insights were easy to read at a glance": 4 somewhat disagree, 1 neither, 4 somewhat agree, 7 strongly agree. "Insights helped advance the conversation": 2 neither, 3 somewhat agree, 11 strongly agree. "Helped discover unexpected information": 2 neither, 2 somewhat agree, 12 strongly agree. "Supported decisions without overriding agency": 1 somewhat disagree, 6 somewhat agree, 9 strongly agree. "Would use this system in future meetings": 1 somewhat disagree, 3 neither, 4 somewhat agree, 8 strongly agree. Across all seven statements, most responses cluster in the somewhat agree and strongly agree categories, with the strongest agreement for helping discover unexpected information and advancing the conversation, and comparatively more mixed responses for insights appearing at the right moments and being easy to read at a glance.}

	\label{fig:system-ratings} 
\end{figure}

All participants were able to complete both tasks. We performed a thematic analysis of the sessions{, organizing our findings into three themes that interleave qualitative with quantitative results shown in} Figure~\ref{fig:results}, Figure~\ref{fig:exp1-task}{, and Figure~\ref{fig:system-ratings}}, following the `weaving' mixed-methods reporting methodology \cite{McChesney2019-of}.

\bstart{T1: Proactive and serendipitous information retrieval results in less task switching}
As reflected in the significantly higher ratings for perceived natural flow and engagement in the conversation (Figure~\ref{fig:results}.A),
participants consistently reported that a proactive delivery of content allowed them to stay more present in the discourse, rather than diverting attention to manual information retrieval. 
This sense of flow can be attributed to the insights surfaced in the side channel, with participants rating them favorably across measures of timeliness,
glanceability {(empirically supporting the text insight length design choice described in Section~\ref{sec:interface})},
and their ability to help advance the conversation {(Figure~\ref{fig:system-ratings})}.
P15 described this as \textit{``a big game changer\ldots\ very subtle and very seamlessly providing me insights,''} adding that they appeared \textit{``at moments of our conversation where it felt appropriate\ldots\ whenever there's ambiguity in the conversation, or when more context is needed.''} 

Some participants described how the {\raisebox{-0.09em}{\inlineicon{0.8em}{figures/magnifier.pdf}} \textit{BL}}{} condition forced them into a dual-task mode, simultaneously conversing and searching~[P04, P13, P16], with P13 noting: \textit{``you can't do efficient discussion, thinking, note taking, and being in a normal mental state\ldots\ but this} [InsightToast]\textit{, you have it in front of you.''} 
In the {\inlineicon{0.75em}{figures/logo.pdf} \textit{\textit{IT}}}{} condition, participants offloaded the search process and reported lower physical demand ($M{=}1.75$ vs.\ $M{=}2.31$; $p{=}.030$),
mentioning that \textit{``it was doing the hard part of searching for information for me, and actually use legit sources''} [P07]. 
This pivot in search behavior is evident in fewer reactive searches 
(Figure~\ref{fig:exp1-task}.B), with P01 stating that \textit{``I didn't have to think of a question to ask''} with proactive retrieval.

In the {\raisebox{-0.09em}{\inlineicon{0.8em}{figures/magnifier.pdf}} \textit{BL}}{} condition, the struggle of search began even before typing a query, with the participants consistently identifying crafting a search term as a central barrier to effective information retrieval, illustrated by remarks such as \textit{``I'm struggling to find keywords to search''}~[P05], \textit{``am I searching up the right keywords?''}~[P15], \textit{``I don't know how to even ask} [chatbot] \textit{and formulate my prompt''}~[P06]. 
Compounding this was the challenge of deciding \emph{which} source would be of most help between the chatbot, Google, and the knowledge base search.
Some participants resolved to only query the chatbot~[P07, P08, P09, P14], albeit with noted reservations about trust, as P06 put it: \textit{``My justification is just the} [chatbot]. \textit{I don't trust that, but I have to go with this.''} 
This cumulative cost of finding information in the baseline was reflected in significantly higher effort 
and frustration (Figure~\ref{fig:results}.B), 
often making participants abandon or delay search altogether, leading to a \textit{``huge break in the continuity of the conversation.''} [P12].
In contrast, P09 noted that in the {\inlineicon{0.75em}{figures/logo.pdf} \textit{\textit{IT}}}{} condition, \textit{``we can actually discuss it in the meeting, and just not postpone it to the next meeting, after I've done the research.''}

Beyond changing post-meeting search behavior, proactive delivery also reshaped how participants spent their time within a session. 
As illustrated by Figure~\ref{fig:exp1-task}.A, a pattern emerged across both conditions: the first ${\sim}60\%$ of session time was dominated by exploration, while the remaining ${\sim}40\%$ shifted toward deliberation, with these phases manifesting differently across conditions. 
In the exploration phase, behavior in the {\raisebox{-0.09em}{\inlineicon{0.8em}{figures/magnifier.pdf}} \textit{BL}}{} condition followed a \emph{pause--search--resume} cycle in which participants repeatedly halted the conversation to formulate queries and scan documents, producing \textit{``long gaps''} [P16], whereas in the {\inlineicon{0.75em}{figures/logo.pdf} \textit{\textit{IT}}}{} condition, exploration happened mostly through interactions with proactively synthesized insights,
which P16 described as allowing ideas to \textit{``build layer on layer.''} 
The deliberation phase also differed, with participants largely moving off the main collaboration interface to read opened document tabs in the {\raisebox{-0.09em}{\inlineicon{0.8em}{figures/magnifier.pdf}} \textit{BL}}{} condition: Figure~\ref{fig:exp1-task}.A-bottom shows long intervals containing no interaction with the system.
In the {\inlineicon{0.75em}{figures/logo.pdf} \textit{\textit{IT}}}{} condition, participants remained within the conversation interface and engaged with citations directly. 

\bstart{T2: Deliberation broadens and decisions feel more informed}
Participants in the {\inlineicon{0.75em}{figures/logo.pdf} \textit{\textit{IT}}}{} condition reported feeling significantly more informed about their decisions ($M{=}4.06$ vs.\ $M{=}2.62$; $p{=}.004$; Figure~\ref{fig:results}.B).
P13 contrasted the two experiences, describing how the \textit{``first decision} [{\raisebox{-0.09em}{\inlineicon{0.8em}{figures/magnifier.pdf}} \textit{BL}}{}] \textit{was mostly based on my perception and my thoughts, not really grounded \ldots\, while the second decision [{\inlineicon{0.75em}{figures/logo.pdf} \textit{\textit{IT}}}{}] was explained by external studies or sources of laws.''} 
To examine whether this subjective sense of being more informed translated to observable differences, we assessed participants' post-task written justifications along three axes inspired by the Toulmin Argument Model~\cite{Toulmin2003-ij}: \emph{claims} (distinct arguments made), \emph{evidence} (factual support for claims), and \emph{citations} (references to specific sources).
To complete this assessment, one member of the research team along with two LLM-as-judges \cite{Zheng2023-fg, Gao2024-go} (GPT-5.4 and Claude Opus 4.6) coded participants' justifications with a high degrees of inter-rater reliability (Krippendorff's $\alpha{=}0.861$).
We observed that the number of claims
($M{=}3.50$ vs.\ $M{=}3.88$)
and evidence pieces 
($M{=}2.50$ vs.\ $M{=}1.94$)
did not differ significantly between conditions, indicating that participants  constructed a similar number of arguments backed by a comparable amount of factual support. 
However, participants in the {\inlineicon{0.75em}{figures/logo.pdf} \textit{\textit{IT}}}{} condition included significantly more source citations in their justifications ($M{=}2.31$ vs.\ $M{=}1.25$; $p{=}.006$), suggesting that they found proactively-retrieved sources relevant enough to cite in support of their arguments.
P02 qualifies this observation by stating: \textit{``I felt like I could cite more to say why I thought certain things. Whereas in the first one} [{\raisebox{-0.09em}{\inlineicon{0.8em}{figures/magnifier.pdf}} \textit{BL}}{}]\textit{, I felt more like I was just saying things that I believed, but I couldn't back it up.''} 
This source traceability corresponded with feeling significantly more informed, with
P07 remarking how they felt \textit{``more informed with the AI system} [{\inlineicon{0.75em}{figures/logo.pdf} \textit{\textit{IT}}}{}]\textit{, specifically because I had more sources to back my point.''} 
The comparable volume of evidence across conditions appears to have provided sufficient justification for participants' reasoning, yielding no significant difference in decision confidence.
Ignorance, it would seem, was bliss in the {\raisebox{-0.09em}{\inlineicon{0.8em}{figures/magnifier.pdf}} \textit{BL}}{} condition, where participants could construct plausible arguments and feel reasonably confident.

Participants in both conditions came to the conversation with pre-existing views on the policy topics, and several described how, without \textsc{InsightToast}, they would have searched only for information confirming their existing position. 
P09 explained, \textit{``I was kind of biased to the other position\ldots\ I believe I would ask} [chatbot] \textit{questions more toward that position, and do not even think about the other position,''} and P12, who described themselves as \textit{``a pretty opinionated person,''} noted that \textit{``the insights were like a good, factually grounded way for me to consider other opinions,''}.
P01 spoke to the root of this mitigation, stating that they were getting \textit{``more perspectives''} because \textit{``it} [InsightToast] \textit{told me about both.''} 
This broadening of perspectives was also reflected in questionnaire results, 
($M=4.44$ vs.\ $M=3.81$; $p=.048$)
with participants reporting a markedly greater sense of having considered all important aspects
($M=3.44$ vs.\ $M=2.19$; $p=.003$).
These broader perspectives often surfaced as unexpected surprises, presenting information that participants would not have thought to seek on their own. 
P04, for instance, described encountering a specific policy detail that was \textit{``not something I would have even considered,''} noting that it \textit{``made me completely recontextualize my thoughts,''} while P07 reported shifting positions entirely after encountering unanticipated evidence, reflecting that without it, they would \textit{``probably have dug my heels into position B.''} 
Crucially, participants perceived this broadening as expanding rather than constraining their agency {(Figure~\ref{fig:system-ratings})},
with P06 valuing that the system \textit{``was not giving me a direct answer, but it was supporting my thoughts.''} 
This sense of support over substitution also fostered trust, particularly through grounded citations, as P12, who described having \textit{``very low trust towards AI systems in general,''} noted that \textit{``the fact that it generated these insights and gave me direct} [policy] \textit{sources, I felt like I was able to trust those insights.''}

\bstart{T3: Adapting to proactive meeting intelligence elicits new usage scenarios}
Participants reported significantly higher enjoyment in the {\inlineicon{0.75em}{figures/logo.pdf} \textit{\textit{IT}}}{} condition
(Figure~\ref{fig:results}.C), and rated the interface favorably 
across its evaluation dimensions.
Participants specifically valued the titled summaries that enabled quick scanning, mentioning that \textit{``when I was searching, I could just skim based on titles,''} [P07], and the clickable citations that provided transparent sourcing, as they could \textit{``click the button and immediately see what it's talking''} [P02].
Reactions to chart-based insights, however, were mixed, with participants who described themselves as "visual learners"~[P06, P13, P15] finding them most useful, stating that \textit{``the charts were really helpful for me to understand,''} [P06] and appreciating that they \textit{``helped really quickly visualize stats,''} [P04], while some \textit{``preferred the text insight generations over the charts, because they were smaller and digestible''} [P07], consistent with prior findings on the efficiency of text for information extraction~\cite{Lee2021-yi}.
Participants who described how demanding gathering data could be in their current meeting practices on the pre-study questionnaire were particularly enthusiastic about proactive meeting intelligence.
Correlating responses from the pre-study questionnaire against participant's gain from {\raisebox{-0.09em}{\inlineicon{0.8em}{figures/magnifier.pdf}} \textit{BL}}{} to {\inlineicon{0.75em}{figures/logo.pdf} \textit{\textit{IT}}}{} conditions, we found significant positive correlations with perceived advantages in information retrieval and decision-making (Spearman $\rho{=}0.50$, $p{=}.049$) and engagement ($\rho{=}0.61$, $p{=}.012$), as well as a trending reduction in workload ($\rho{=}{-}0.47$, $p{=}.069$), indicating that \textsc{InsightToast} especially benefits those with high search overhead in their current practices.

Participants were enthusiastic about the prospect of using applications like \textsc{InsightToast} in future meetings {(Figure~\ref{fig:system-ratings})}.
Even P07, who expressed concerns about AI's environmental impact, acknowledged that \textit{``this seems like a really good place''} for AI, as it is \textit{``not really replacing the creative aspect of being a human''} but rather \textit{``replacing some of the grunt work.''} 
Participants connected the system to diverse knowledge bases suited to their own contexts, including academic research literature~[P05, P07, P12], cross-domain collaboration where knowledge gaps are known to exist~[P06], policy analysis with government archives~[P15, P16], and even casual settings like vacation planning~[P02].
Notably, P10 described a scenario of missing institutional context within an organization: after senior team members depart and juniors \textit{``reinvent the wheel''} because they are unaware of past activities documented in an organizational knowledge base; they envision \textsc{InsightToast} as a bridge to this institutional knowledge.

Participants also described how they adapted to \textsc{InsightToast}, with P08 stating \textit{``it's like getting a phone for the first time,''} while P05 observed it was \textit{``a lot to get started with, but just like any software, I think I will get comfortable using it,''} mentioning \textsc{InsightToast}'s novelty effect. 
This feeling of adaptation is notable given the information density of the 15-minute task durations across study sessions, with an average of $28.53$ insights ($SD{=}8.58$) generated over each session ($\approx$ one insight every 30s). 
However, this density of proactively delivered content also introduced its own disruptions, with P04 describing how their \textit{``train of thought was hijacked''} by insight toasts, P03 noting they \textit{``had to sift through a lot more noise compared to manually searching,''} and P08 pointing to \textit{``the cognitive load of having a conversation while also deciphering what these insights meant.''}
Beyond adapting \emph{to} the system, participants also envisioned how the system could adapt \emph{to} them. 
A recurring desire was reactive insight generation alongside the proactive mechanism, using proactivity for breadth while allowing targeted depth on demand: P03 wanted \textit{``proactive themes, but the insights that I'm looking for within the theme, I would want more agency \ldots\ to manually do it.''} 
However, P04 indicated that reactive insight generation does not necessarily mean chatbot interaction, but rather \textit{``a button in the transcript where I could select} [an utterance] \textit{and say, generate me insights on this thing.''} 
Several participants also identified the potential value for longer meeting sessions, where they expected the topic tracker to be more useful~[P03, P08, P11], and in meetings with more participants, where one can read insights while others speak~[P04, P05]. 
Beyond using \textsc{InsightToast} during conversations, participants also saw value in retrospective engagement. P04, who found real-time interaction overwhelming, would instead \textit{``turn it on, let it run,} [and] \textit{review it after the meeting,''} fully deferring engagement with synthesized content to post-meeting, while P11 would \textit{``dive deeper into them} [insights] \textit{later''}, only skimming during the meeting. P10 suggested pinning as a mechanism for more efficient use of insights after the meeting and envisioned a \textit{``conversation after the conversation with the} [retrieved] \textit{documents.''}

\subsection{Technical Evaluation}
\label{sec:technical-eval}

To assess the performance of \textsc{InsightToast}'s pipeline, we analyzed  processing logs from all 16 study sessions along three dimensions:
\textit{(i) processing latency:} Each 15-minute session processed an average of 113 transcribed utterances ($SD{=}18$), with a mean end-to-end pipeline latency of $7.5s$ per chunk ($SD{=}13.8$\,s). Utterances that triggered insight synthesis required substantially more processing ($M{=}29.1$\,s, $SD{=}22.0$\,s) compared to those that resolved without synthesis ($M{=}3.4$\,s, $SD{=}5.3$\,s).
\textit{(ii) LLM inference cost:} Each session consumed approximately $2.17$M input ($SD{=}0.54$M) and $141$K output ($SD{=}42$K) tokens across all pipeline agents. At the time of writing, Gemini Flash, used for chart insight agents, is priced at \$0.50/\$3.00 per 1M input/output tokens, while Gemini Flash Lite, used for the remaining pipeline agents (see Section~\ref{sec:implementation}), is priced at \$0.10/\$0.40, resulting in a mean LLM inference cost of \$0.45 per session.
\textit{(iii) proactive retrieval quality:} To assess the accuracy of knowledge gap detection that triggers retrieval, a member of the research team annotated whether each utterance across all sessions warranted retrieval. We compared these labels against the system's binary classification, yielding a mean $F1=0.81$ (precision${=}0.70$, recall${=}0.99$), indicating the system rarely missed genuine information needs, with occasional over-triggering at adjacent turns.



\section{Discussion}
\label{sec:discussion}

We reflect on \textsc{InsightToast}'s design and evaluation, the limitations of our research, and future work in proactive meeting intelligence.

\bstart{Reflections}
Although proactive retrieval reduced task switching {by delegating query formulation and evidence synthesis to the system and reducing the temporal and spatial separation between communication and information discovery}, some participants characterized the frequency of toast notifications as \textit{noise}. We attribute this to occasional over-triggering in the pipeline and high information density in the session. 
{While this relocation of task switching was unfamiliar early in the study sessions,} participants appeared to grow more comfortable identifying relevant content as they adapted to the interface. {Rather than relying on such adaptation alone,} a more granular model of notification criticality that goes beyond knowledge-gap detection could inform more intelligent interruption decisions \cite{Adamczyk2004-vi}. 
By attaching levels of criticality to toasts, accounting for conversational state and task progress, the system could allow participants to tune the notification threshold through meeting settings, balancing the exploration and exploitation \cite{Berger-Tal2014-st} for the meeting's information needs.
For instance, participants could opt to mute new notifications during retrospective meeting review.

Beyond calibrating \textit{when} insights surface, our findings also point to refining \textit{what} triggers their generation. 
While ambient discourse currently serves as the exclusive input modality initiating insight synthesis {and implicitly capturing participant feedback}, direct voice commands, gestural cues \cite{Bolt1980-aq}, and inline UI controls could add reactive channels for on-demand generation {and for reporting failures such as irrelevant or erroneous insights}, shifting toward higher levels of control in the mixed-initiative design spectrum \cite{Riley1989-hj}. 
Our study participants valued insights not merely for being proactive but for their context-aware and digestible form that eliminated the burden of formulating search queries, suggesting that reactive mechanisms could preserve this quality while offering targeted depth on demand. 
Context detection could also be refined by attending to deictic expressions and referential cues in speech \cite{Srinivasan2023-zi}, such as demonstrative references to entities or concepts under discussion, using these signals alongside speech utterances to better scope what the system retrieves. {More broadly, while \textsc{InsightToast} currently establishes its meeting and knowledge base context through explicit configuration in its codebase and implicit inference from its running memory of the discourse, richer pre-meeting specifications, including agendas and a priori meeting goals~\cite{Chen2025-lg}, could further steer both retrieval and synthesis.}

\bstart{Limitations}
On the technical side, the mean synthesis latency of 29 seconds per insight-triggering utterance risks brief conversational disruptions as participants pause to await the results. Additionally, the inference cost, measured at approximately half a dollar per fifteen-minute session, may limit broader adoption. Adapting open-weight models hosted on internal servers within adopting organizations presents one path forward, though this introduces a quality-cost trade-off, as smaller models may degrade the pipeline's quality, while also addressing within-organization data privacy requirements, pointing to \textsc{InsightToast}'s potential adaptability to domains involving sensitive data. 
{Moreover, as with any system reliant on LLMs, bias and hallucination persist even under retrieval grounding~\cite{Sharma2024-tp, Huang2025-hi}. While \textsc{InsightToast} tempers such risks through multi-faceted retrieval and traceable citations that invite verification, disclosing this fallibility to participants alongside adopting more reliable models can serve as additional safeguards.}

Methodologically, our evaluation involved trade-offs for a within-subjects contrast within a single study session. While this setup provided an initial demonstration of \textsc{InsightToast}'s capabilities, it does not capture the dynamics of larger groups deliberating over longer periods with multiple subject-matter experts.

\bstart{Future Work}
Our methodological limitations call for further evaluation with varying group sizes, conversation topics, and meeting durations.
Data-rich collaborative tasks beyond decision-making are also exciting directions to consider~\cite{Brehmer2026-jo}, such as group brainstorming and real-time data monitoring.
Beyond limitations, the extension of applications like \textsc{InsightToast} into specialized application domains suggests new opportunities for proactive meeting information retrieval.
For instance, conversations that are concurrent with collaborative interaction around maps and Geographic Information System (GIS) interfaces could support discussions among urban planners.
Similarly, multimodal discussion and interaction with medical imagery and electronic health records (EHRs) could surface clinically-relevant insights.
In either setting, the delivery of notifications may need to be adapted to appear adjacent to or near the object, image, or map location being manipulated, and the palette of chart insights will likely need to be expanded beyond quantitative and temporal attributes.
{Beyond charts, other insight modalities merit further investigation in accordance with their application contexts. In meeting settings, while our findings supported the glanceability of \textsc{InsightToast}'s tweet-length snippet cap, a focused study of text length under the constrained conversational bandwidth of live discourse could better inform the design of informative yet non-disruptive of insights in text insights.}

{More explicit configurability could better align proactive support with participants' desired degree of control over serendipitous information discovery. Promising dimensions include tunable proactivity levels, alternative interruption and triggering mechanisms, and assignable meeting roles, as participants who differ in data access privileges and domain knowledge may hold asymmetric information needs for maintaining common ground.} We envision a system like \textsc{InsightToast}, adaptable across meeting modalities and generalizable through the substitution of its indexed knowledge base, being adopted across diverse settings, from corporate boardrooms to academic advisory meetings, where surfacing the right data at the right moment is critical.



\section{Conclusion}

We contributed an approach to proactive information delivery for synchronous meetings, one with the aim of filling gaps in knowledge to ground the conversation in data.
We encapsulated our approach in a meeting application called \textsc{InsightToast}, suitable for both co-located and remote/hybrid meetings; as its name implies, it delivers ephemeral toast notifications in participants' periphery at opportune moments in a conversation.
These notifications are (ideally) insightful, summarizing content primarily from an institutional knowledge base and secondarily from the open web, in the form of either pithy text statements or glanceable charts, with dual attribution to the conversation and to source documents.
We evaluated \textsc{InsightToast} in a study ($N=16$) where we contrasted the experience of using it in a meeting emulating a legislative decision-making scenario with a baseline reactive search interface. 
Our findings suggest that participants were quite receptive to proactive meeting assistance, especially for identifying \textit{unknown unknowns} and serendipitously surfacing content that would otherwise have gone amiss. 
Ultimately, this work is an initial step toward generalized proactive support for the collaborative analysis of heterogeneous data types and the integration of this support into the longitudinal fabric of collaborative knowledge work.



\begin{acks}
{We thank S. Onay, D. Vogel, A. Crisan, S. Amirshahi, and members of the CS HCI Lab and UBIX research group at U. Waterloo. }
\end{acks}

\bibliographystyle{ACM-Reference-Format}




\appendix

\clearpage

\renewcommand\thefigure{\thesection.\arabic{figure}}
\renewcommand\thetable{\thesection.\arabic{table}}
\setcounter{figure}{0}
\setcounter{table}{0}


\section{Appendix: Pipeline Implementation Details}
\label{app:pipeline-details}

{This appendix details the proactive retrieval-trigger and chart-design criteria referenced in Section~\ref{sec:pipeline}.}

\subsection{Knowledge Gap Detection Criteria}
\label{app:gap-criteria}

{Retrieval is triggered when the current utterance, read against the recent conversational window and the meeting's running context (Section~\ref{sec:pipeline}), satisfies at least one of the conditions in Table~\ref{tab:gap-triggers}; for pragmatic reasons, topical relevance or an isolated hedge word alone does not qualify. Categories are disjunctive while each category's own conditions remain conjunctive, weighting the resulting judgment toward recall over precision (Section~\ref{sec:technical-eval}), since a dismissable toast is a lower-cost outcome than a missed gap in a high-stakes deliberation. Detection is thus aimed at resolving ambiguity and preventing discussion from stalling, opening room for participants to pursue avenues they might not otherwise raise.}

\begin{table*}[t]
\centering
\caption{{Trigger conditions evaluated by the knowledge gap detection agent. Retrieval is initiated only when all conditions within at least one trigger category are satisfied.}}
\label{tab:gap-triggers}
\small
\begin{tabular}{@{}p{3.2cm}p{14.2cm}@{}}
\toprule
\textbf{Trigger} & \textbf{Condition} \\
\midrule
\multirow{3}{3.2cm}{Explicit knowledge gap}
  & A speaker explicitly requests or admits lacking specific background information, past data, or historical context \\
  & The gap is concrete and specific, not vague or rhetorical \\
  & The gap cannot be resolved from information already available within the conversation \\
\midrule
\multirow{3}{3.2cm}{Factual or historical inquiry}
  & A direct question is posed about past events, decisions, precedents, or official records \\
  & The speaker indicates that obtaining this information would help the discussion proceed \\
  & The question cannot be answered from within the conversation itself \\
\midrule
\multirow{2}{3.2cm}{Contested or uncertain claims}
  & A verifiable claim or assumption is potentially incorrect or actively contested \\
  & The uncertainty is visibly affecting the direction of the discussion \\
\midrule
\multirow{2}{3.2cm}{Evidence-dependent decision}
  & A concrete, unresolved question is on the table that participants are actively trying to resolve \\
  & Resolving it requires factual evidence not currently available to any participant \\
\midrule
\multirow{2}{3.2cm}{Stalled discussion}
  & The discussion has become circular, with no forward progress across recent turns \\
  & Participants lack the factual grounding needed to advance \\
\bottomrule
\end{tabular}
\end{table*}

\subsection{Chart Design Constraints and Verification}
\label{app:chart-constraints}

{Chart type selection follows the analytic intent and data shape of the retrieved evidence, drawn from the restricted inventory in Table~\ref{tab:chart-inventory}.} {Suppression of non-essential elements (Section~\ref{sec:interface}) is enforced as fixed generation rules rather than case-by-case judgment: legends are added only after simplifying categories fails to resolve ambiguity, gridlines appear only when they aid reading a quantitative axis, and exact values are exposed via tooltip rather than on-chart labels. The generation agent verifies the resulting specification against these rules before returning it, regenerating it if validation fails.}

\begin{table}[H]
\centering
\caption{{Restricted chart type inventory used during visualization planning, organized by data shape and analytic intent. Secondary types require explicit justification that no primary type serves the intent.}}
\label{tab:chart-inventory}
\footnotesize
\begin{tabular}{@{}p{2.1cm}p{5.95cm}@{}}
\toprule
\textbf{Data Shape (Intent)} & \textbf{Chart Types} \\
\midrule
Category $\times$ Value \newline \textit{(Comparison/Ranking)} & Simple Bar, Horizontal Bar, Sorted Bar \newline \textit{Secondary:} Grouped Bar, Horizontal Grouped Bar, Bar w/ Highlighted Bar \\
\midrule
Category $\times$ Value \newline \textit{(Composition)} & Stacked Bar, Horizontal Stacked Bar, Normalized Stacked Bar, Horizontal Normalized Stacked Bar \newline \textit{Secondary:} Diverging Stacked Bar \\
\midrule
Time $\times$ Value \newline \textit{(Trend)} & Simple Line, Multi-Series Line, Step \newline \textit{Secondary:} Stacked Area, Slope Graph, Bump Chart \\
\midrule
Parts-of-Whole \newline \textit{($\leq$6 categories)} & Donut, Pie \\
\midrule
Event Distribution & Strip Plot \\
\midrule
Grid/Matrix & \textit{Secondary:} Heatmap \\
\bottomrule
\end{tabular}
\end{table}

\section{Appendix: Evaluation Task Materials}

{This appendix documents the materials used to administer the deliberation task described in Section~\ref{sec:evaluation}.}

\subsection{Task Briefs}
\label{app:study-task}

As described in Section~\ref{sec:evaluation}, each task was framed as a citizen advisory brief in which participants recommended one of two competing policy positions. We gave each participants a task card presenting the policy question, both positions with a brief description and arguments for and against to better introduce different aspects of the positions, and three justification criteria to consider during deliberation: \textit{value for public money} (whether the approach is a responsible use of government funding), \textit{evidence of effectiveness} (whether the approach has been studied or discussed by government), and \textit{political feasibility} (whether there is evidence of prior legislative support or opposition). {The task cards below were presented to participants, one per condition, each framing a policy decision between two competing positions.}

\vspace{1em}

\noindent\textbf{Task 1: Rental Housing Affordability}

\vspace{0.5em}

\noindent\textit{How should the federal government address rental housing affordability for citizens?}

\vspace{-0.5em}

\aptLtoX{\definecolor{TFFrameColor}{rgb}{0.831,0.443,0.29}
\definecolor{TFTitleColor}{rgb}{1,1,1}
\begin{quote}
{\setlength{\fboxsep}{6pt}\colorbox{TFFrameColor}{\textcolor{TFTitleColor}{\textbf{\textbf{Position A:} Introduce a national rent increase cap}}}}

Limits how much landlords can raise rent each year, tied to the rate of inflation (CPI).

\vspace{4pt}
\textbf{For:} Protects existing tenants from sudden large rent increases while longer-term solutions are developed.

\vspace{2pt}
\textbf{Against:} Critics argue it discourages landlords from building new rental units, shrinking supply over time.%
\end{quote}%
\begin{quote}
{\setlength{\fboxsep}{6pt}\colorbox{TFFrameColor}{\textcolor{TFTitleColor}{\textbf{\textbf{Position B:} Fund purpose-built rental construction}}}}

Provides federal tax incentives and direct funding to developers who build new rental-only housing.

\vspace{4pt}
\textbf{For:} Increases the number of available rental units, addressing the root cause of high prices.

\vspace{2pt}
\textbf{Against:} Critics argue the benefits are long-term and do nothing for renters struggling with costs right now.%
\end{quote}}{\noindent
\begin{minipage}[t]{0.48\textwidth}
\taskcard{deeppeach}{\textbf{Position A:} Introduce a national rent increase cap}{%
Limits how much landlords can raise rent each year, tied to the rate of inflation (CPI).

\vspace{4pt}
\textbf{For:} Protects existing tenants from sudden large rent increases while longer-term solutions are developed.

\vspace{2pt}
\textbf{Against:} Critics argue it discourages landlords from building new rental units, shrinking supply over time.%
}
\end{minipage}%
\hfill
\begin{minipage}[t]{0.48\textwidth}
\taskcard{deeppeach}{\textbf{Position B:} Fund purpose-built rental construction}{%
Provides federal tax incentives and direct funding to developers who build new rental-only housing.

\vspace{4pt}
\textbf{For:} Increases the number of available rental units, addressing the root cause of high prices.

\vspace{2pt}
\textbf{Against:} Critics argue the benefits are long-term and do nothing for renters struggling with costs right now.%
}
\end{minipage}}

\vspace{1.5em}

\noindent\textbf{Task 2: Healthcare Wait Times}

\vspace{0.5em}

\noindent\textit{What should the federal government do to reduce healthcare wait times for citizens?}

\vspace{-0.5em}

\aptLtoX{\definecolor{TFFrameColor}{rgb}{0.431,0.392,0.769}
\definecolor{TFTitleColor}{rgb}{1,1,1}
\begin{quote}
{\setlength{\fboxsep}{6pt}\colorbox{TFFrameColor}{\textcolor{TFTitleColor}{\textbf{\textbf{Position A:} Conditional Health Funding}}}}

Makes a portion of federal health funding to provinces/states conditional on meeting and publicly reporting specific wait time targets.

\vspace{4pt}
\textbf{For:} Creates measurable accountability for how provinces/states spend federal health dollars.

\vspace{2pt}
\textbf{Against:} Critics argue it penalizes provinces/states where chronic underfunding and staff shortages are already severe.%
\end{quote}%
\begin{quote}
{\setlength{\fboxsep}{6pt}\colorbox{TFFrameColor}{\textcolor{TFTitleColor}{\textbf{\textbf{Position B:} Federal hospital staffing \& infrastructure investment}}}}

Provides direct federal funding to hospitals and health authorities specifically to hire clinical staff.

\vspace{4pt}
\textbf{For:} Directly addresses the staffing shortages and capacity gaps driving long wait times.

\vspace{2pt}
\textbf{Against:} Critics argue that without conditions attached, provinces/states have no obligation to maintain improvements after funding ends.
\end{quote}}{\noindent
\begin{minipage}[t]{0.48\textwidth}
\taskcard{deeppurple}{\textbf{Position A:} Conditional Health Funding}{%
Makes a portion of federal health funding to provinces/states conditional on meeting and publicly reporting specific wait time targets.

\vspace{4pt}
\textbf{For:} Creates measurable accountability for how provinces/states spend federal health dollars.

\vspace{2pt}
\textbf{Against:} Critics argue it penalizes provinces/states where chronic underfunding and staff shortages are already severe.%
}
\end{minipage}%
\hfill
\begin{minipage}[t]{0.48\textwidth}
\taskcard{deeppurple}{\textbf{Position B:} Federal hospital staffing \& infrastructure investment}{%
Provides direct federal funding to hospitals and health authorities specifically to hire clinical staff.

\vspace{4pt}
\textbf{For:} Directly addresses the staffing shortages and capacity gaps driving long wait times.

\vspace{2pt}
\textbf{Against:} Critics argue that without conditions attached, provinces/states have no obligation to maintain improvements after funding ends.}
\end{minipage}}

\vspace{1em}

{The full study instruments, including the pre- and post-task questionnaires, the closing semi-structured interview guide, and the original task cards as presented to participants are provided as supplemental material.}


\subsection{Conversation Partner Cues}
\label{app:task-script}

{During each deliberation, a member of the research team served as the participant's conversation partner, using the scripted cues below to build rapport, sustain the conversation, and keep the discussion on topic without steering participants toward either position. We organized cues into three phases matching the deliberation stage the participant was in, and bracketed placeholders (\eg~[topic]) indicate the specific pair of positions assigned to each task.}

\noindent\textbf{Introduction}
\begin{itemize}
    \item {``Both sound reasonable, but reading them, what is your initial reaction to them?''}
    \item {``That makes sense. But I'm wondering: do we actually know if Parliament has looked at this seriously before? Has there been any committee work on [topic] specifically?''}
    \item {``What were the findings on similar approaches before?''}
\end{itemize}

\noindent\textbf{Knowledge Base Exploration}
\begin{itemize}
    \item {``I wonder if there's been a committee study on this, with actual recommendations, not just debate.''}
    \item {``The card mentions [an argument]. Has Parliament actually heard from [experts/committee researchers] on that? I'd want to know what the evidence says before we take that argument at face value.''}
    \item {``Has any bill on [this problem] actually made it through Parliament, or do they all die on the order paper?''}
    \item {``What do the voting records look like on [bills supporting this]? Is this something with cross-party support or is it completely partisan?''}
    \item {``There must be petitions on [this problem]. I wonder how much public pressure Parliament has actually received on this.''}
\end{itemize}

\noindent\textbf{Deliberation}
\begin{itemize}
    \item {``OK, so based on what we've found, which of these positions actually holds up on the value-for-money side?''}
    \item {``Is there enough political will in Parliament for either of these to actually pass? Because a great policy that goes nowhere doesn't help anyone.''}
    \item {``Even if we're not fully certain, which position do we feel we can defend better with what we found?''}
\end{itemize}



\end{document}